\documentclass{ar-1col}
\usepackage[utf8]{inputenc}
\usepackage{url}
\usepackage{natbib}
\usepackage{overpic,rotating}
\usepackage{wrapfig}

\jname{Xxxx. Xxx. Xxx. Xxx.}
\jvol{AA}
\jyear{YYYY}
\doi{10.1146/((please add article doi))}

\usepackage[centertags]{amsmath}
\usepackage{amsfonts,amscd,amssymb}

\usepackage{epstopdf}
\usepackage{latexsym}
\usepackage{graphicx}
\graphicspath{{figures/}}

\usepackage{xparse}
\usepackage{tikz}
\usepackage{bm}
\usetikzlibrary{shapes,arrows}
\tikzstyle{block} = [draw, fill=blue!20, rectangle, minimum height=3em, minimum width=3em]
\tikzstyle{sum} = [draw, fill=blue!20, circle, node distance=1cm]
\tikzstyle{input} = [coordinate]
\tikzstyle{output} = [coordinate]
\tikzstyle{pinstyle} = [pin edge={to-,thin,black}]
\tikzstyle{pinstyle2} = [pin edge={to-,thin,black}]

\usepackage{color}
\definecolor{piblue}{RGB}{0, 10, 122}

\definecolor{maroon}{rgb}{.8,0,0}

\begin{document}
% Page header
\markboth{Brunton, Noack, and Koumoutsakos}{Machine Learning for Fluid Mechanics}

% Title
\title{Machine Learning for Fluid Mechanics}

%Authors, affiliations address.
\author{Steven L. Brunton,$^1$ Bernd R. Noack$^2$ $^3$ and Petros Koumoutsakos$^4$ 
\affil{$^1$Mechanical Engineering, University of Washington, Seattle, WA, USA, 98195}
\affil{$^2$ LIMSI, CNRS, Universit\'e Paris-Saclay, F-91403 Orsay, France}
\affil{$^3$ Institut f\"ur Str\"omungsmechanik und Technische Akustik, TU Berlin, D-10634, Germany}
\affil{$^4$ Professorship for Computational Science, ETH Zurich, CH-8092, Switzerland; email: petros@ethz.ch}}

%Abstract
\begin{abstract}
The field of fluid mechanics is rapidly advancing, driven by unprecedented volumes of data from field measurements, experiments and large-scale simulations at multiple spatiotemporal scales.  
Machine learning offers a wealth of techniques to extract information from data that could be translated into knowledge about the underlying fluid mechanics. 
Moreover, machine learning algorithms can augment domain knowledge and automate tasks related to flow control and optimization.  
This article presents an overview of past history, current developments, and emerging opportunities of machine learning for fluid mechanics.  
It outlines fundamental machine learning methodologies and discusses their uses for understanding, modeling, optimizing, and controlling fluid flows.  
The strengths and limitations of these methods are addressed from the perspective of scientific inquiry that considers data as an inherent part of  modeling, experimentation, and simulation.
Machine learning provides a powerful information processing framework that can enrich, and possibly even transform, current lines of fluid mechanics research and industrial applications.

\end{abstract}

%Keywords, etc.
\maketitle

%Table of Contents
% \tableofcontents

\newpage

%% SECTION IN ALL CAPS
%% Subsection All First Letter Caps
%% Subsubsections only have first word capitalized
%%%%%%%%%%%%%%%%%%%%%%
%%% INTRODUCTION
%%%%%%%%%%%%%%%%%%%%%%
\section{INTRODUCTION}\label{sec:intro}
Fluid mechanics has traditionally dealt with massive amounts of data from experiments, field measurements, and large-scale numerical simulations.  \emph{Big data} has been a reality in fluid mechanics~\citep{Pollard2016book} over the last decade due to high-performance computing architectures and advances in experimental measurement capabilities. 
Over the past 50 years many techniques were developed to handle data of fluid flows, ranging from advanced algorithms for data processing and compression, to databases of turbulent flow fields~\citep{perlman2007data,wu2008direct}. However, the analysis of fluid mechanics data has relied ,to a large extent, on domain expertise, statistical analysis, and heuristic algorithms. % and mostly  linear algorithms. 

\begin{figure}
    \centering
    \vspace{-.35in}
    \includegraphics[width=1.175\textwidth]{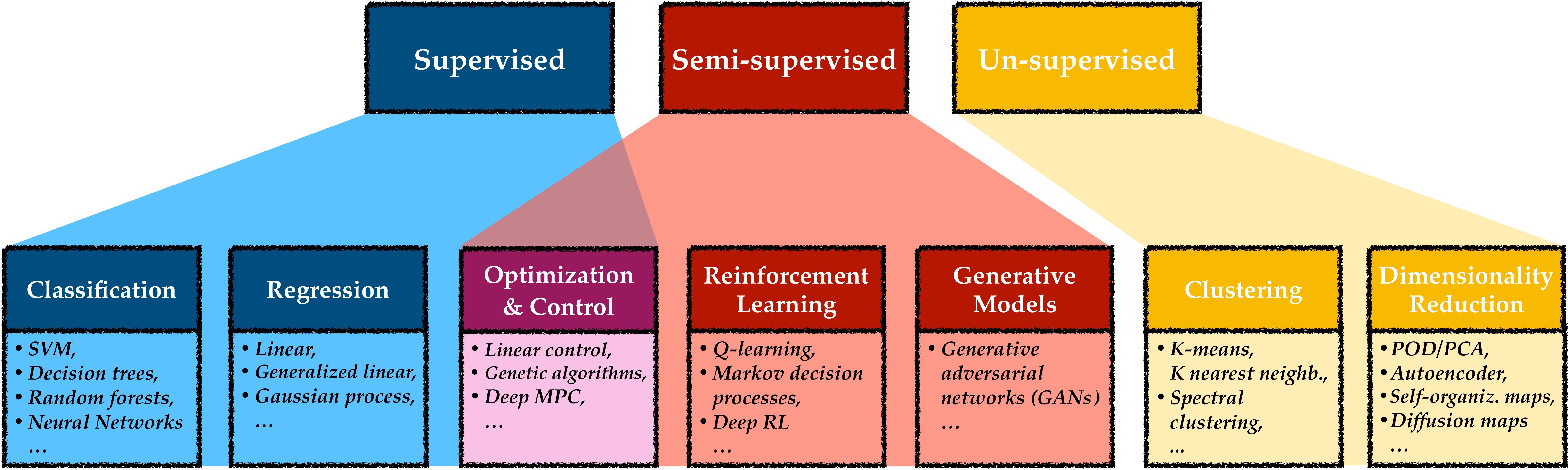}
    \vspace{-0.1in}
    \caption{Machine learning algorithms may be categorized into supervised, unsupervised, and semi-supervised, depending on the extent and type of  information available for the learning process.}
    \label{fig:FlowSchematic}
    \vspace{-0.15in}
\end{figure}

Massive amounts of data is today widespread across scientific disciplines, and gaining insight and actionable information from them  has become a new mode of scientific inquiry as well as a commercial opportunity. 
Our generation is experiencing an unprecedented confluence of 1) vast and increasing volumes of data, 2) advances in computational hardware and reduced costs for computation, data storage and transfer, 3) powerful algorithms, 4) an abundance of open source software and benchmark problems, and 5) significant and ongoing investment by industry on data driven problem solving. 
These advances have, in turn, fueled renewed interest and progress in the field of machine learning (ML).  Machine learning algorithms (here categorized as supervised, semi-supervised, and unsupervised learning (see Fig.~\ref{fig:FlowSchematic}) are  rapidly making inroads in fluid mechanics. 
Machine learning provides a modular and agile modeling framework that can be tailored to address many challenges in fluid mechanics, such as reduced-order modeling, experimental data processing, shape optimization, turbulence closure, and control. As scientific inquiry increasingly shifts from first principles to data-driven approaches, we  may draw a parallel between current efforts in machine learning with the development of numerical methods in the 1940's and 1950's to solve the equations of fluid dynamics.  
Fluid mechanics stands to benefit from learning algorithms and in return present challenges that may further advance these algorithms to complement human understanding and engineering intuition.  

\begin{marginnote}[]
\entry{Machine learning}{Algorithms that extract patterns and information from data. They facilitate automation and can augment human domain knowledge.}
%\entry{Optimization}{The algorithmic search for the ``best" values that maximize or minimize a given objective function within an allowable set of values.}
%\entry{Regression}{A statistical model that represents an outcome or output variable in terms of input or indicator variables.}
\end{marginnote}

In this review, in addition to outlining successes, we emphasize the importance of understanding  how learning algorithms work and when these methods succeed or fail. 
It is important to balance excitement about the capabilities of machine learning with the reality that its application to fluid mechanics is an open and challenging field.  
In this context, we also highlight the benefit of incorporating domain knowledge about fluid mechanics into learning algorithms. 
We envision that the fluid mechanics community can contribute to advances in machine learning reminiscent of the advances in numerical methods in the last century.

\subsection{Historical Overview}\label{Sec:Intro:History}

Machine learning and fluid dynamics share a long, and possibly surprising, history of interfaces. 
In the early 1940's Kolmogorov, a founder of statistical learning theory, considered turbulence as one of its prime application domains~\citep{Kolmogorov1941}. Advances in machine learning in the 1950's and 1960's were characterized by two distinct developments. On one side we distinguish cybernetics~\citep{Wiener1965} and  expert systems  designed to emulate the thinking process of the human brain, and on the other ``machines" like the perceptron~\citep{rosenblatt1958perceptron} aimed to automate processes such as classification and regression. 
Advances on the second branch are also prevailing today and it is understandable how the use of perceptrons for classification created significant excitement for Artificial Intelligence (AI) in the early 50's.
However, this excitement was quenched by findings that their capabilities had severe limitations~\citep{minsky1969perceptrons}: single layer  perceptrons were only able to learn linearly separable functions and not capable of learning the XOR function. It was known that multi-layer perceptrons could learn the XOR function, but perhaps their advancement was limited given the computational resources of the times (a recurring theme in Machine Learning research).
The reduced interest in perceptrons was soon accompanied by a reduced interest in AI in general. 

\begin{marginnote}[-10pt]
\entry{Perceptron}{The first learning machine: A composition of binary decision units used for classification.}
\end{marginnote}

\begin{figure}
    \centering
    \vspace{-.4in}
    \includegraphics[width=1.00\textwidth]{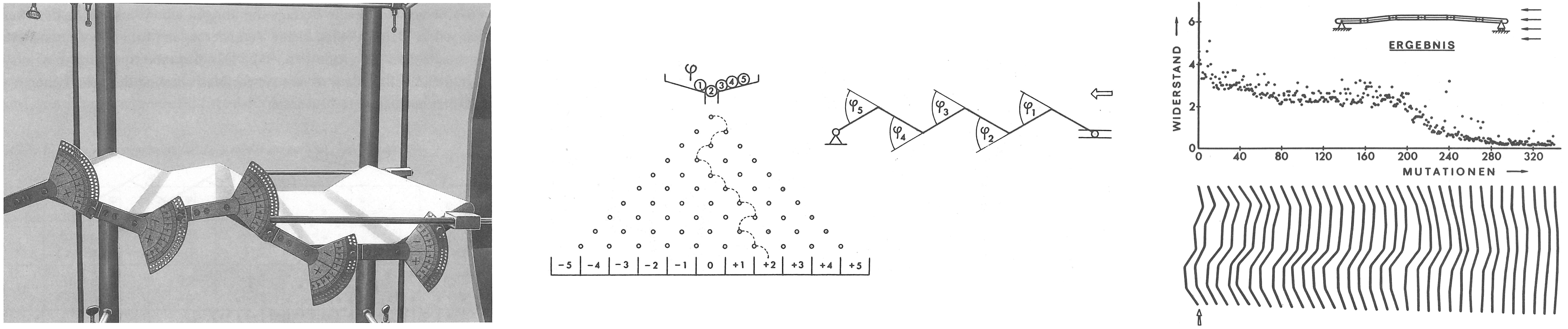}
    \vspace{-.1in}    
    \caption{First example of learning and automation in experimental fluid mechanics: Rechenberg's experiments for optimally corrugated plates for drag reduction using the Galtonbrett (Galton board) as an analog random number generator~\citep{rechenberg1964kybernetische}. }  
    %Presented in the joint meeting of the  WGLR and  DGRR on September 16, 1964 in the Berlin  Kongresshalle with the title: ``Kybernetische L\"osungsansteuerung einer experimentellen Forschungsaufgabe"
    \label{fig:RechenbergExp}
        \vspace{-.2in}    
\end{figure}

Another branch of machine learning, closely related to the budding ideas of cybernetics in the early 1960's, was pioneered by two graduate students: Ingo Rechenberg and Hans-Paul Schwefel at TU Berlin. 
They performed experiments in a wind tunnel on a corrugated structure composed of 5 linked plates with the goal of finding their optimal angles to reduce the overall drag (see Fig.~\ref{fig:RechenbergExp}).
Their breakthrough involved adding random variations to these angles, where the randomness was generated using a Galton board (an ``analog" random number generator).  Most importantly, the size of the variance was learned (increased/decreased) based on the success rate (positive/negative) of the experiments. 
The work of Rechenberg and Schwefel has received little recognition, even though over the last decades a significant number of applications in fluid mechanics and aerodynamics use ideas that can be traced back to their work.
Renewed interest in the potential of AI for aerodynamics applications materialized almost simultaneously with the early developments in computational fluid dynamics in the early 1980's. Attention was given to expert systems to assist in aerodynamic design and development processes~\citep{Mehta1984}.

An indirect link between fluid mechanics and machine learning was the so-called ``Lighthill report" in 1974 that criticized artificial intelligence programs in the UK, as not delivering on their grand claims.  This report played a major role in the reduced funding and interest in AI in the UK and subsequently in the USA, known as the \emph{AI winter}. 
Lighthill's main argument was based on his perception that AI would never be able  to address the challenge of the combinatorial explosion between possible configurations in the parameter space. He used the limitations of language processing systems of that time as a key demonstration of that failure for AI. In Lighthill's defense, 40 years ago the powers of modern computers as we know them today may have been difficult to fathom. Indeed today one may watch Lighthill's speech against AI on the internet while a machine learning algorithm automatically provides the captions.

The reawakening of interest in machine learning, and in neural networks in particular, came in the late 1980's with the development of the back-propagation algorithm~\citep{rumelhart1988learning}. 
This enabled the training of neural networks with multiple layers, even though in the early days at most two layers were the norm. 
Another source of stimulus were the works by~\cite{hopfield1982neural,gardner1988space,hinton1986learning} who developed links between machine learning algorithms and statistical mechanics. 
However, these developments did not attract many researchers from fluid mechanics. 

In the early 1990's a  number of applications of neural networks in flow-related problems were developed in the context of trajectory analysis and classification for particle tracking velocimetry (PTV) and particle image velocimetry (PIV) ~\citep{TeoNNPIV91,grant1995investigation} as well as to identify phase configurations in multi-phase flows~\citep{Bishop1993}. 
The link between POD and linear neural networks~\citep{baldi1989neural} was exploited in order to reconstruct turbulence flow fields and the flow in the near wall region of a channel flow using wall only information ~\citep{Milano2002jcp}. 
This application introduced multiple layers of neurons to improve compression results, marking perhaps the first use of deep learning, as it is known today, in the field of fluid mechanics. 

In the past few years we have experienced a renewed blossoming of machine learning applications in fluid mechanics. 
Much of this interest is attributed to the remarkable performance of deep learning architectures, which hierarchically extract informative features from data. 
This has led to several advances in data rich and model limited fields, such as social sciences, and in companies for which prediction is a key financial factor. 
Fluid mechanics is not model-limited and is rapidly becoming a data rich field. We believe that this confluence of first principles and data-driven approaches is unique and has the potential to transform both fluid mechanics and machine learning.

\begin{textbox}[b]
\section{LEARNING FLUID MECHANICS: FROM  LIVING ORGANISMS TO MACHINES}
% %%%%%%%%%%%%%%%%%%%%%%%%%%%%%%%%
Birds, bats, insects, fish, and other aquatic and aerial lifeforms, perform remarkable feats of fluid manipulation. They optimize and control their shape and motion to harness unsteady fluid forces for agile propulsion, efficient migration, and other maneuvers.  
The range of fluid optimization and control observed in biology has inspired humans for millennia.  How do these organisms \emph{learn} to manipulate the flow environment?

To date, we know of only one species that manipulates fluids through knowledge of the Navier-Stokes equations. Humans have been innovating and engineering devices to harness fluids since before the dawn of recorded history, from dams and irrigation, to mills and sailing.  
Early efforts were achieved through intuitive design, although recent quantitative analysis and physics-based design have enabled a revolution in performance over the past hundred years.  
Indeed, physics-based engineering of fluid systems is a high-water mark of human achievement.  
However, there are serious challenges associated with equation-based analysis of fluids, including high-dimensionality and nonlinearity, which defy closed-form solutions and limit real-time optimization and control efforts.
At the beginning of a new millennium, with increasingly powerful tools in machine learning and data-driven optimization, we are again learning how to learn from experience.  
\end{textbox}

%%%%%%%%%%%%%%%%%%%%%%
%%% CHALLENGES IN FLUID DYNAMICS
%%%%%%%%%%%%%%%%%%%%%%
\subsection{Challenges and Opportunities for Machine Learning in Fluid Dynamics}\label{sec:challenges}
Fluid dynamics presents challenges that differ from those tackled in many applications of machine learning, such as image recognition and advertising.  
In fluid flows it is often important to precisely quantify the underlying physical mechanisms in order to analyze them. 
Furthermore, fluids flows entail complex, multi-scale phenomena whose understanding and control remain to a large extent unresolved. 
Unsteady flow fields require algorithms capable of addressing nonlinearities and multiple spatiotemporal scales that may not be present in popular machine learning algorithms. 
In addition, many prominent applications of machine learning, such as playing video games, rely on inexpensive system evaluations and an exhaustive categorization of the process that must be learned.  
This is not the case in fluids, where experiments may be difficult to repeat or automate and where simulations may require large-scale supercomputers operating for extended periods of time.  

Machine learning has also become instrumental in robotics, and algorithms such as reinforcement learning are used routinely in autonomous driving and flight. 
While many robots operate in fluids, it appears that the subtleties of fluid dynamics are not presently  a major concern in their design. Reminiscent of the pioneering days of flight, solutions imitating natural forms and processes are often the norm (see the sidebar titled "Learning Fluid Mechanics: From Living Organisms to Machines").
We believe, that the deeper understanding and exploitation of fluid mechanics will become critical in the design of robotic devices, when their energy consumption and reliability in complex flow environments become a concern.

%%%%%%%%%%%%%%%%%%%%%%%%%%%%%%%%%%%%%%%%%%%%%%%%%%%%
\begin{marginnote}
\entry{Interpretability}{The degree to which a model may be understood or interpreted by an expert human.}
\entry{Generalizability}{The ability of a model to generalize to new examples including unseen data.  Newton's second law is an example.}
\end{marginnote}
%%%%%%%%%%%%%%%%%%%%%%%%%%%%%%%%%%%%%%%%%%%%%%%%%%%%

In the context of flow control, actively or passively manipulating flow dynamics for an engineering objective may change the nature of the system, making predictions, based on data of uncontrolled systems, impossible. 
Although fluid data is vast in some dimensions, such as spatial resolution, it may be sparse in others; e.g., it may be expensive to perform parametric studies.  Furthermore, fluids data can be highly heterogeneous, requiring special care when choosing the type of learning machine. 
In addition, many fluid systems are non-stationary, and even for stationary flows it may be prohibitively expensive to obtain statistically converged results.

Fluid dynamics are central to transportation, health, and defense systems, and it is, therefore, essential that machine learning solutions are interpretable, explainable, and generalizable. 
Moreover, it is often necessary to provide guarantees on performance, which are presently  rare. 
Indeed, there is a poignant lack of convergence results, analysis, and guarantees in many machine learning algorithms.   
It is also important to consider whether the model will be used for interpolation within a parameter regime or for extrapolation, which is considerably more challenging. 
Finally, we emphasize the importance of \emph{cross-validation} on  withheld data sets to prevent overfitting in machine learning. 

We suggest that this, non-exhaustive, list of challenges need not be a barrier; to the contrary, it should provide a strong motivation for the development of more effective machine learning techniques.  These techniques will likely impact a number of  disciplines if they are able to solve fluid mechanics problems. 
For example, the application of machine learning to systems with known physics, such as fluid mechanics, may provide deeper theoretical insights into the effectiveness of these algorithms. 
We also believe that hybrid methods, combining  machine learning and first principles models, will be a fertile ground for development.

 This review is structured as follows:  Section 2 outlines the fundamental algorithms of machine learning, followed by their applications to flow modeling (Sec. 3), and optimization and control (Sec. 4). We provide a summary and outlook of this field in Sec. 5. 

%%%%%%%%%%%%%%%%%%%%%%
%%% MACHINE LEARNING
%%%%%%%%%%%%%%%%%%%%%%
\section{MACHINE LEARNING FUNDAMENTALS} \label{sec:machinelearning}

\begin{figure}
    \centering
    % \vspace{-.2in}
    \includegraphics[width=.61\textwidth]{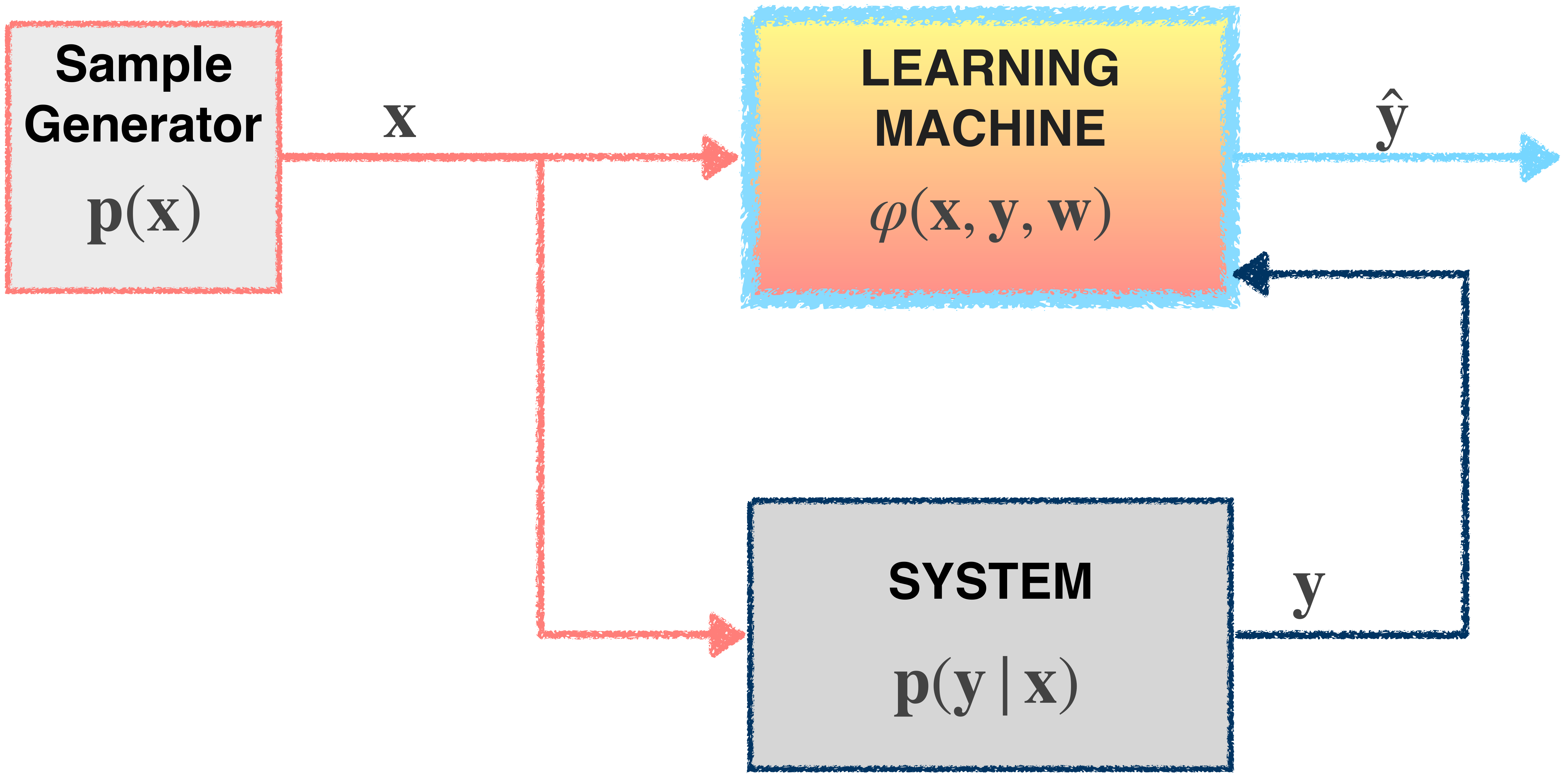}
    \caption{The learning problem: A learning machine uses inputs from a sample generator and observations from a system to generate an approximation of its output (Credit: \cite{Cherkasskybook}). Symbols:
  $x$ — inputs
  $p(x)$ — probability distribution of inputs
  $y$ — outputs
  $\hat{y}$ — estimated outputs, given input $x$
  $w$ — parameters of the learning machine
  $\varphi(x,y,w)$ — structure of the learning machine
  $p(y|x)$ — probability of outputs given input $x$}
    \label{fig:Learning} 
\end{figure}

The learning problem can be formulated as the process of estimating associations between inputs, outputs, and parameters of a system using a limited number of observations~\citep{Cherkasskybook}. 
We distinguish a generator of samples, the system in question, and a learning machine (LM), as in Fig.~\ref{fig:Learning}. We emphasize that the approximations by learning machines are fundamentally stochastic and their learning process can be summarized as the minimization of  a risk functional:
\begin{equation}\label{Eq:RiskFunctional}
    R({\bf w}) \;=\; \int L({\bf y}, {\boldsymbol{\phi}}({\bf x},{\bf y},{\bf w}))\;p({\bf x},{\bf y})\,d{\bf x}d{\bf y},
\end{equation}
where  the data ${\bf x}$ (inputs) and ${\bf y}$ (outputs) are samples from a probability distribution $p$, ${\boldsymbol{\phi}}({\bf x},{\bf y},{\bf w})$ defines the structure and ${\bf w}$ the parameters of the learning machine, and the loss function $L$ balances the various learning objectives (e.g., accuracy, simplicity, smoothness, etc.). 
We emphasize that the risk functional is weighted by a probability distribution $p({\bf x,y})$ that also constrains the predictive capabilities of the learning machine. %, to inputs from the same probability distribution.
The various types of learning algorithms can be grouped into three major categories: Supervised, unsupervised and semi-supervised, as in Fig.~\ref{fig:FlowSchematic}. 
These distinctions signify the degree to which external supervisory information from an expert is available to the learning machine.
\begin{marginnote}[60pt]
\entry{Supervised learning}{Learning from labeled data by providing corrective information to the algorithm.}
\entry{Unsupervised learning}{Learning patterns (e.g clusters, classes) without labeled training data.} 
\entry{Semi-supervised learning}{Learning with partially labeled data (GANs) or through receiving a reward from the environment (Reinforcement learning).}
\end{marginnote}

\subsection{Supervised Learning}

Supervised learning implies the availability of corrective information to the learning machine. In its simplest and most common form, this implies labeled training data, with labels corresponding to the output of the LM. 
Minimization of the cost function, which implicitly depends on the training data, will determine the unknown parameters of the LM. 
In this context, supervised learning dates back to the regression and interpolation methods proposed centuries ago by  Gauss~\citep{Meijering2002}. 
A commonly employed loss function is \begin{equation}
    L({\bf y}, {\boldsymbol{\phi}}({\bf x},{\bf y},{\bf w})) = ||{\bf y}- {\boldsymbol{\phi}}({\bf x},{\bf y},{\bf w})||^2.
\end{equation} 
Alternative loss functions may reflect different constraints on the learning machine such as sparsity~\citep{Hastie2009book,Brunton2019book}. The choice of the approximation function reflects prior knowledge about the data and the choice between linear and  nonlinear methods directly bears on the computational cost associated with the learning methods. 

\subsubsection{Neural networks}

Neural networks are arguably the most well known methods in supervised learning. 
They are fundamental nonlinear function approximators, and in recent years a number of efforts have been dedicated in understanding their  effectiveness. 
The universal approximation theorem~\citep{hornik1989multilayer} states that any function may be approximated by a sufficiently large and deep network.  Recent work has  shown that sparsely connected, deep neural networks are information theoretic optimal nonlinear approximators for a wide range of functions and systems~\citep{Bolcskei2019}.  
\begin{marginnote}
\entry{Neural network}{A computational architecture, based loosely on biological networks of neurons. Neural networks are often used for nonlinear regression.  
\hspace*{0in}\includegraphics[width=.225\textwidth]{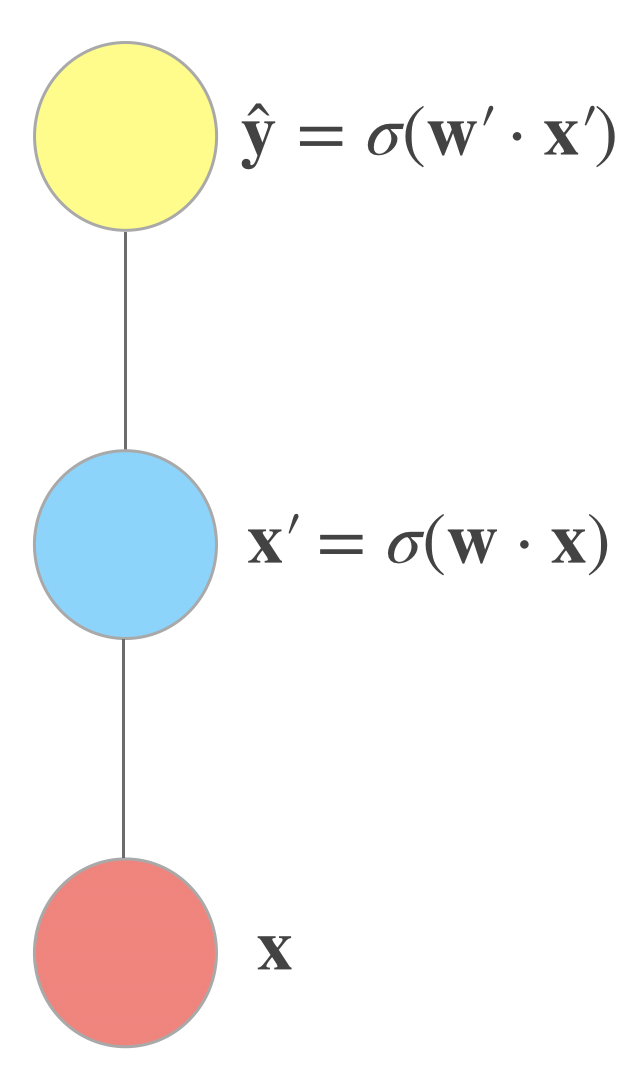}}
\end{marginnote}
%A simple neural network with input ${\bf x}$, output $\hat{\bf y}$, activation function $\sigma$ and weights ${\bf w, w'}$ that are  determined from data ${\bf y}$ by minimizing ${\bf E}=|| {\bf y} - \hat{\bf y} ||^2$.

The power and flexibility of neural networks emanates from their modular structure based on the neuron as a central building element, a caricature of the neurons in the human brain. 
Each neuron receives an input, processes it through an activation function, and produces an output. 
Multiple neurons can be combined into different structures that reflect knowledge about the problem and the type of data. 
Feed-forward networks are among the most common structures, and they are composed of layers of neurons, where a weighted output from one layer is the input to the next layer.
NN architectures have an input layer that receives the data and an output layer that produces a prediction. 
Nonlinear optimization methods, such as back-propagation~\citep{rumelhart1988learning}, are used to identify the network weights to minimize the error between the prediction and labeled training data.  
Deep neural networks involve multiple layers and various types of nonlinear activation functions. 
When the activation functions are expressed in terms of convolutional kernels, a powerful class of networks emerges, namely convolutional neural networks (CNN), with great success in image and pattern recognition~\citep{Krizhevsky2012nips,Goodfellow-et-al-2016}.

\begin{figure}
\centering
\vspace{-.2in}
\includegraphics[trim=15 0 15 0,clip,width=1.\textwidth]{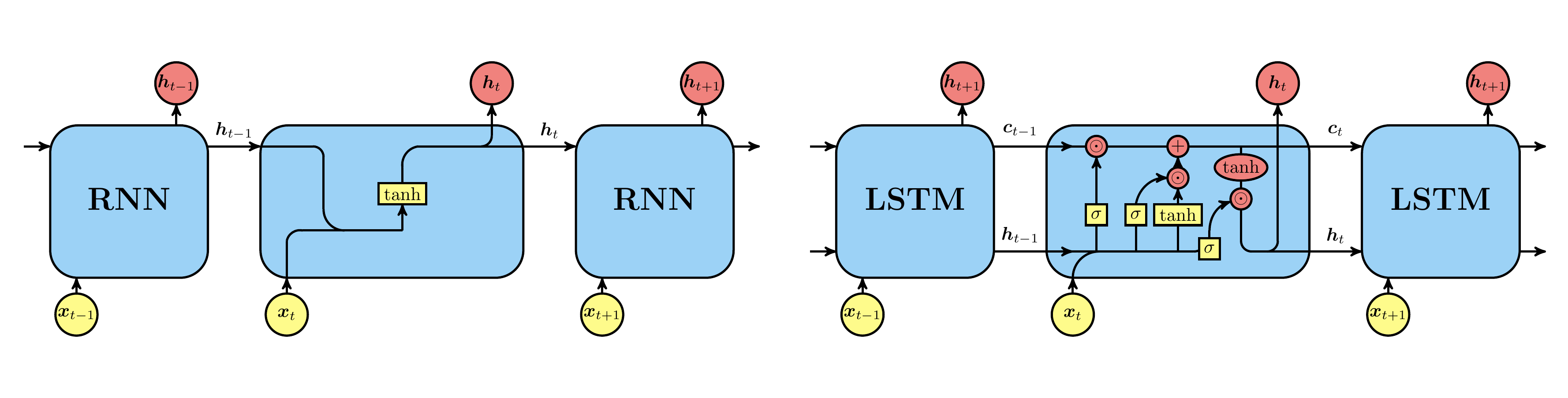}
\vspace{-.2in}
\caption{Recurrent neural nets (RRNs) for time-series predictions and the Long Short-Term Memory (LSTM) regularization (\cite{hochreiter1997long}). Symbols: $h_{t-1}$ — previous cell output
  $h_{t}$ — current cell output
  $x_{t}$ — input vector
  $c_{t-1}$ — previous cell memory
  $c_{t}$ — current cell memory
  $\sigma$ — sigmoid }
\label{fig:RNNLSTM}
\vspace{-.1in}
\end{figure}

Recurrent neural networks (RNNs), depicted in Fig.~\ref{fig:RNNLSTM}, are of particular interest to fluid mechanics. 
They operate on sequences of data (e.g., images from a video, time-series, etc.) and their weights are obtained by back-propagation through time (BPTT). 
RNNs have been quite successful for natural language processing and speech recognition. Their architecture  takes into account the inherent order of the data, thus augmenting some of the pioneering applications of classical neural networks on signal processing \citep{rico-martinez1992}
However, the effectiveness of RNNs has been hindered by diminishing or exploding gradients that emerge during their training. 
The renewed interest in RNNs is largely attributed to the development of the long short-term memory (LSTM)~\citep{hochreiter1997long} algorithms that deploy cell states and gating mechanisms to store and forget information about past inputs, thus alleviating the problems with gradients and the transmission of long-term information that standard RNNs suffer from. 
An extended architecture, called the multi-dimensional LSTM network (MD-LSTM)~\citep{Graves2007}, was proposed to efficiently handle high-dimensional spatiotemporal data. 
A number of potent alternatives to RNNS have appeared over the years; notably the echo state networks have been used with success to predict certain  dynamical systems~\citep{pathak2018model}. 

\subsubsection{Classification: Support vector machines and random forests}

Classification is a supervised learning task that can determine the label or category of a set of measurements from a-priori labeled training data. 
It is perhaps the oldest method for learning, starting with the perceptron~\citep{rosenblatt1958perceptron}, which could classify between two types of linearly separable data. 
Two fundamental classification algorithms are support vector machines (SVM)~\citep{Scholkopf2002book} and random forests~\citep{Breiman2001ml}, which have been widely adopted in the industry for several learning tasks, until the recent progress by deep neural networks.  
The problem can be specified by a loss functional, which is most simply expressed for two classes: 

\begin{equation}
 L \big({\bf y}, {\boldsymbol{\phi}}({\bf x},{\bf y},{\bf w}) \big) = 
    \begin{cases}
    0, & \text{if ${\bf y}={\boldsymbol{\phi}}({\bf x},{\bf y},{\bf w})$},\\
    1, & \text{if ${\bf y}\neq{\boldsymbol{\phi}}({\bf x},{\bf y},{\bf w})$}.
  \end{cases} 
\end{equation} 
Here the output of the learning machine is an indicator on the class to which the data belong.
The risk functional quantifies the probability of misclassification and the task is to minimize the risk based on the training data by suitable choice of ${\boldsymbol{\phi}}({\bf x},{\bf y},{\bf w})$.
Random forests are based on an ensemble of decision trees that hierarchically split the data using simple conditional statements; these decisions are interpretable and fast to evaluate at scale.  
In the context of classification, an SVM maps the data into a high-dimensional  feature space on which a linear classification is possible. 
%%% SOMETHING LOOKS WRONG IN THE SVM LOSS FUNCTION... HOW CAN Y MULTIPLY PHI?
% The SVM loss function can be described as  
% \begin{equation}
%  L_\Delta \big( {\bf y}, {\boldsymbol{\phi}}({\bf x},{\bf y},{\bf w}) \big) = \max \big( \Delta - {\bf y}{\boldsymbol{\phi}}({\bf x},{\bf y},{\bf w}),0 \big) 
% \end{equation}
% where $\Delta$ defines a margin for the classification error. 

%%%%%%%%%%%%%%%%%%%%%%%%%%%%%%%%%%%%%%%%%%%%%%%%%%%%%%%%%%%%%%%%%%%%

\begin{marginnote}
\entry{Deep learning}{Neural networks with multiple interconnected layers that can create hierarchical representations of the data. }
\end{marginnote}
%%%%%%%%%%%%%%%%%%%%%%%%%%%%%%%%%%%%%%%%%%%%%%%%%%%%%%%%%%%%%%%%%%%%

%%%%%%%%%%%%%%%%%%%%%%%%%%%%%%%%%%%%%%%%%%%%%%%
\subsection{Unsupervised Learning}
This learning task implies the extraction of features from the data by specifying certain global criteria and without the need for supervision or a ground-truth label for the results. 
The types of problems involved here include dimensionality reduction, quantization, and clustering. 
The automated extraction of flow features by unsupervised learning algorithms can form the basis of flow modeling and control using low-order models.

% \begin{figure}
%      \centering
%      % The following vspace messes up with previous figures
%      %\vspace{-.35in}
%      \hspace{-.1in}
%      \begin{overpic}[width=1.22\textwidth]{NNautoencoder3.pdf}
%      \put(27,0.){${\bf x}$}
% \put(45.8,0.){${\bf \hat{x}}$}
% \put(52.9,0.){${\bf x}$}
% \put(120.5,0.){${\bf \hat{x}}$}
% \put(36.8,9.){${\bf z}$}
% \put(86.7,10){${\bf z}$}
% \put(78.9,24.){$\boldsymbol{\varphi}({\bf x})$}
% \put(90.9,24.){$\boldsymbol{\psi}({\bf z})$}
% \put(32.4,24){${\bf U}$}
% \put(40.4,24){${\bf V}$}
% \end{overpic}
%      \vspace{-.1in}
%   \caption{PCA/POD (left) vs shallow autoencoder (sAE, middle), versus deep autoencoder (dAE, right).  If the node activation functions in sAE are linear, then ${\bf U}$ and ${\bf V}$ are matrices that minimize the loss function $\| {\bf \hat{x}} - {\bf V U x}\|$.  The node activation functions may be nonlinear,  minimizing the loss function $\| {\bf x} - {\boldsymbol{\psi} \big( \boldsymbol{\varphi}({\bf x}) \big)}\|$.  
%   The input  ${\bf x}\in\mathbb{R}^D$ is reduced to ${\bf z}\in\mathbb{R}^M$, with $M\ll D$. Note that the PCA/POD requires the solution of a problem-specific eigenvalue equation. The neuron modules can be extended to nonlinear activation functions and multiple nodes and layers. (adapted from C. Bishop)}
%     \label{NNautoencoder}
%     \vspace{-.1in}
% \end{figure}
\begin{figure}
     \centering
     \vspace{-.25in}\hspace{-.1in}
     \begin{overpic}[width=1.22\textwidth]{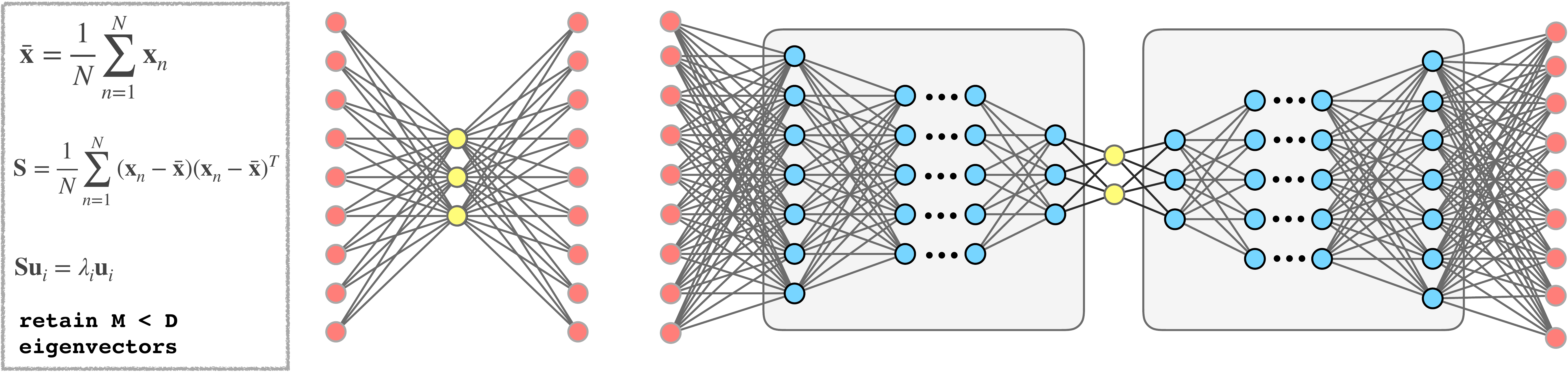}
\put(27.,0.){${\bf x}$}
\put(45.5,0.){${\bf \hat{x}}$}
\put(52.5,0.){${\bf x}$}
\put(120.5,0.){${\bf \hat{x}}$}
\put(36.3,9.){${\bf z}$}
\put(86.5,10){${\bf z}$}
\put(79.3,24.){$\boldsymbol{\varphi}({\bf x})$}
\put(90.5,24.){$\boldsymbol{\psi}({\bf z})$}
\put(32.5,24){${\bf U}$}
\put(39.5,24){${\bf V}$}
\end{overpic}
     \vspace{-.1in}
  \caption{PCA/POD (left) vs shallow autoencoder (sAE, middle), versus deep autoencoder (dAE, right).  If the node activation functions in sAE are linear, then ${\bf U}$ and ${\bf V}$ are matrices that minimize the loss function $\| {\bf \hat{x}} - {\bf V U x}\|$.  The node activation functions may be nonlinear,  minimizing the loss function $\| {\bf x} - {\boldsymbol{\psi}( \boldsymbol{\varphi}({\bf x}))}\|$.  
  The input  ${\bf x}\in\mathbb{R}^D$ is reduced to ${\bf z}\in\mathbb{R}^M$, with $M\ll D$. Note that the PCA/POD requires the solution of a problem specific eigenvalue equation while the neuron modules and can be  extended to nonlinear activation functions and multiple nodes and layers (adapted from \cite{Bishop1993}). Symbols: $x_n$ — $n$-th input vector
  $\bar{x}$ — mean of input data
 $ S$ — covariance matrix of mean-subtracted data
  $u_i$ — eigenvector
  $\lambda_i$ — eigenvalue
  $x$ — input vector
  $\hat{x}$ — autoencoder reconstruction
  $\varphi(x)$ — deep encoder
  $\psi(x)$ — deep decoder
  $U$ — linear encoder
  $V$ — linear decoder
  $z$ — latent variable}
    \label{NNautoencoder}
\end{figure}

\subsubsection{Dimensionality reduction I : POD, PCA and auto-encoders}
The extraction of flow features from experimental data and large scale simulations is a cornerstone for flow modeling.
Moreover identifying lower dimensional representations for high-dimensional data can be used as pre-processing for all tasks in supervised learning algorithms.
Dimensionality reduction can also be viewed as an ``information filtering bottleneck" where the data is processed through a lower dimensional representation before being mapped backed to the ambient dimension.
The classical proper orthogonal decomposition (POD) algorithm belongs to this category of learning, and will be discussed more in Sec.~\ref{sec:modeling}. 
The POD, or linear principal components analysis (PCA) as it is more widely known, can be formulated as a two layer neural network (an autoencoder) with a linear activation function for its linearly weighted input, that can be trained by stochastic gradient descent (see Fig.~\ref{NNautoencoder}).
This formulation is an algorithmic alternative to linear eigenvalue/eigenvector problems in terms of neural networks, and it offers a direct route to the nonlinear regime and deep learning by adding more layers and a nonlinear activation function on the network. 
Unsupervised learning algorithms have seen limited use in the fluid mechanics community, and we believe that this is an opportunity that deserves further exploration.
In recent years, the machine learning community has produced numerous auto-encoders that, when properly matched with the possible features of the flow field, can lead to significant insight for reduced-order modeling of stationary and time-dependent data.
%%%%%%%%%%%%%%%%%%%%%%%%%%%%%%%%%%%%%%%%%%%%%%%%%%%%%%%%%%%%%%%%%%%%%%%%%%%%
\begin{marginnote}[]
\entry{Autoencoder}{ A neural network architecture used to compress and decompress high-dimensional data. They are powerful alternatives to the Proper Orthogonal Decomposition (POD).}
\end{marginnote}
%%%%%%%%%%%%%%%%%%%%%%%%%%%%%%%%%%%%%%%%%%%%%%%%%%%%%%%%%%%%%%%%%%%%%%%%%%%%%

\subsubsection{Dimensionality reduction II: Discrete principal curves and self-organizing maps}
The mapping between high-dimensional data and a low-dimensional representation  can be structured through an explicit shaping of  the lower dimensional space, possibly reflecting an a-priori knowledge about this subspace. 
These techniques can be seen as  extensions of the linear auto-encoders, where the encoder and decoder can be nonlinear functions. 
This nonlinearity may come however at the expense of losing the inverse relationship between the encoder and decoder functions that is one of the strengths of linear PCA.
An alternative is to define the decoder as an approximation of the inverse of the encoder, leading to the method of principal curves.
Principal curves are structures on which the data are projected during the encoding step of the learning algorithm. In turn the decoding step amounts to an approximation of the inverse of  this mapping by adding for example some smoothing onto the principal curves.
An important version of this process is the self-organizing map (SOM) introduced by~\cite{KohonenSOM}.
In SOMs the projection subspace is described into a finite set of values with specified connectivity architecture and distance metrics.
The encoder step amounts to identifying for each data point the closest node point on the SOM and the decoder step is a weighted regression estimate, using for example kernel functions, that take advantage of the specified distance metric between the map nodes. 
This modifies the node centers, and the process can be iterated until the empirical risk of the autoencoder has been minimized.
The SOM capabilities can be exemplified by comparing it to linear PCA for two dimensional set of points.
The linear PCA will provide as an approximation the least squares straight line between the points whereas the SOM will map the points onto a curved line that  better approximates the data. 
We note that SOMs can be extended to areas beyond floating point data and they offer an interesting way for creating data bases based on features of flow fields.

\subsubsection{Clustering and vector quantization}
Clustering is an unsupervised learning technique that identifies similar groups in the data. 
The most common algorithm is $k$-means clustering, which partitions data into $k$ clusters; an observation belongs to the cluster with the nearest centroid, resulting in a partition of data space into Voronoi cells.

Vector quantizers identify representative points for data that can be partitioned into a predetermined number of clusters. These points can then be used instead of the full data set so that future samples can be approximated by them. The vector quantizer ${\boldsymbol{\phi}} \big({\bf x},{\bf w} \big)$ provides a mapping between the data ${\bf x}$ and the coordinates of the cluster centers. The loss function is usually the squared distortion of the data from the cluster centers, which must be minimized to identify the parameters of the quantizer:
\begin{equation}
 L({\boldsymbol{\phi}}({\bf x},{\bf w})) = || {\bf x} - {\boldsymbol{\phi}}({\bf x},{\bf w})||^2.
\end{equation}
We note that vector quantization is a data reduction method, not necessarily employed for dimensionality reduction.
In the latter the learning problem seeks to identify low dimensional features in high dimensional data, whereas quantization amounts to finding representative clusters of the data.
Vector quantization must also be distinguished from clustering as in the former the number of desired centers is determined a-priori whereas clustering aims to identify meaningful groupings in the data. When these groupings are represented by some prototypes then clustering and quantization have strong similarities.

%%%%%%%%%%%%%%%%%%%%%%%%%%%%%%%%%%%%%%%%%%%%
\subsection{Semi-Supervised Learning}

Semi-supervised learning algorithms operate under \emph{partial} supervision,
either with limited labeled training data, or with other corrective information from the environment.
Two algorithms in this category are generative adversarial networks (GAN) and  reinforcement learning (RL).
In both cases the learning machine is (self-)trained through a game like process as discussed below. 

\subsubsection{Generative adversarial networks (GAN)}
GANs are learning algorithms that result in a generative model, i.e. a model that produces data according to a probability distribution, which mimics that of the data used for its training.
The learning machine is composed of two networks that compete with each other in a zero sum game~\citep{goodfellow2014generative}.
The generative network produces candidate data examples that are evaluated by the discriminative, or \emph{critic}, network to optimize a certain task.
The generative (G) network's training objective is to synthesize novel examples of data to \emph{fool} the discriminative network into misclassifying them as belonging to the true data distribution.
The weights of these networks (N) are obtained through a process, inspired by game theory, called adversarial (A) learning.
The final objective of the GAN training process is to identify the generative model that produces an output that reflects the underlying system.
Labeled data are provided by the discriminator network and the function to be minimized is the Kullback-Leibler divergence between the two distributions.
In the ensuing ``game", the discriminator aims to maximize the probability of it discriminating between true data and data produced by the generator, while the generator aims to minimize the same probability.
Because the generative and discriminative networks essentially train themselves, after initialization with labeled training data, this procedure is often referred to as \emph{self-supervised}. 
This self-training process adds to the appeal of GANs but at the same time one must be cautious on whether an equilibrium will ever be reached in the above mentioned game.
As with other training algorithms, large amounts of data help the process but, at the moment, there is no guarantee of convergence.

\subsubsection{Reinforcement learning}
Reinforcement learning (RL) is a mathematical framework for problem solving~\citep{Sutton2018book} that implies goal-directed interactions of an agent with its environment. In RL the agent has a repertoire of actions and perceives states. Unlike in supervised learning, the agent does not have labeled information about the correct actions, but instead learns from its own experiences, in the form of rewards that may be infrequent and partial; thus, this is referred to as semi-supervised learning. Moreover, the agent is not concerned only with uncovering patterns in its actions or in the environment, but also with maximizing its long term rewards.
Reinforcement learning is closely linked to dynamic programming~\citep{bellman1952} as it also models interactions with the environment as a Markov decision process.
Unlike dynamic programming, RL does not require a model of the dynamics, such as a Markov transition model, but proceeds by repeated interaction with the environment through trial-and-error. We believe that it is precisely this approximation that makes it highly suitable for complex problems in fluid dynamics.
The two central elements of RL are the agent's policy, a mapping $a = \pi(s)$ between the state $s$ of the system and the optimal action $a$, and the value function $V(s)$ that represents the utility of reaching the state $s$ for maximizing the agent's long-term rewards. 

Games are one of the key applications of RL that exemplify its strengths and limitations. One of the early successes of RL is the backgammon learner of~\cite{TESAURO1992}.
The program started out from scratch as a novice player, trained by playing a couple of million times against itself, won the computer backgammon olympiad, and eventually became comparable to the three best human players in the world.
In recent years, advances in high-performance computing and deep neural-network architectures have produced agents that are capable of performing at or above human performance at video games and strategy games that are much more complicated than backgammon, such as Go~\citep{Mnih2015} and the AI gym~\citep{Mnih2015,silver2016}. 
It is important to emphasize that RL requires significant computational resources due to the large numbers of episodes required to properly account for the interaction of the agent and the environment.
This cost may be trivial for games but it may be prohibitive in experiments and flow simulations, a situation that is rapidly changing~\citep{Verma2018EfficientLearning}. 

A core challenge for RL is the long-term credit assignment (LTCA) problem, especially when rewards are sparse or very delayed in time (for example consider the case of a perching bird or robot).
LTCA implies inference, from a long sequence of states and actions, of causal relations between individual decisions and rewards.
A number of efforts address these issues by augmenting an originally sparsely-rewarded objective with densely-rewarded subgoals~\citep{Schaul2015}.
A related issue is the proper accounting of past experiences by the agent as it actively forms a new policy \cite{Novati2019}.

\subsection{Stochastic Optimization: A Learning Algorithms Perspective}
Optimization is an inherent part of  learning, as a risk functional is minimized in order to identify the parameters of the learning machine. There is, however, one more link that we wish to highlight in this review: optimization (and search) algorithms can be cast in the context of learning algorithms and more specifically as the process of learning the probability distribution of the design points that maximize a certain objective. This connection was pioneered by \cite{Rechenberg:71,Schwefel:74}, who introduced Evolution Strategies (ES) and adapted the variance of their search space based on the success rate of their experiments.  This process is also reminiscent of the operations of selection and mutation that are key ingredients of Genetic Algorithms (GA) \citep{Holland1975book} and Genetic Programming \citep{Koza1992book}. ES and GAs algorithms can be considered as  hybrids between gradient search strategies,
which may effectively march downhill towards a minimum,
and Latin-Hypercube or Monte-Carlo sampling methods, which maximally explore the search space. 
Genetic programming was developed in the late 1980s by J.~R.~Koza, a PhD student of John Holland. 
Genetic programming generalized parameter optimization to function optimization, initially coded as a tree of operations~\citep{Koza1992book}. 
A critical aspect of these algorithms  is that they  rely on an iterative construction of the probability distribution, based on data values of the objective function. This iterative construction can be lenhthy and practically impossible for problems with expensive objective function evaluations.

Over the past twenty years,  ES and GAs  have begun to converge into the framework of estimation of distribution algorithms (EDAs).  The CMA-ES algorithm ~\citep{Ostermeier1994,Hansen2003}  is a prominent example of evolution strategies using an adaptive estimation of the covariance matrix of a Gaussian probability distribution, to guide the search for optimal parameters. This covariance matrix is adapted iteratively  using the best points in each iteration. The CMA-ES is  closely related to a number of other algorithms including the mixed Bayesian optimization algorithms (MBOAs)~\citep{Pelikan:2004}, and the reader is referred to  \cite{Kern:2004} for a comparative review. 
In recent years, this line of work has evolved into the more generalized information-geometric optimization (IGO) framework~\citep{Ollivier2017}. IGO algorithms allow for families of  probability distributions whose parameters are learned during the optimization process and maintain the cost function invariance as a  major design principle. The resulting algorithm makes no assumption on the objective function to be optimized and its flow is equivalent to a stochastic gradient descent. These techniques have been proven to be effective on a number of simplified benchmark problems; however, their scaling remains unclear and there are few guarantees for convergence in cost function landscapes such as those encountered in complex fluid dynamics problems.
We note also that there is interest in deploying these methods in order to minimize the cost functions associated with classical machine learning tasks \citep{Salimans2017}.

\subsection{Important Topics We Have Not Covered: Bayesian Inference, Gaussian Processes}
There are a number of learning algorithms that this review does not address, but which demand particular attention from the fluid mechanics community.
First and foremost we wish to mention {\it Bayesian inference}, which aims to inform the model structure and its parameters from data in a probabilistic framework.
Bayesian inference is fundamental for uncertainty quantification, and it is also fundamentally a learning method, as data are used to adapt the model estimates.
An  alternative view casts machine learning algorithms in a Bayesian framework~\citep{TheodoridisML,Barberbook}. 
The above mentioned optimization algorithms provide also a link between these two views.  Whereas optimization algorithms aim to provide the best parameters of a model for given data in a stochastic manner, Bayesian inference aims to provide the full probability distribution of the model parameters.
It may be argued that Bayesian inference is an even more powerful language than machine learning, as it provides probability distributions for all parameters, leading to robust predictions, rather than single values, as is usually the case with classical machine learning algorithms.
However, a key drawback for Bayesian inference is its computational cost, as it involves sampling and integration in high-dimensional spaces, which can be prohibitive for expensive function evaluations (e.g. wind tunnel experiments or large scale DNS).
Along the same lines one must mention Gaussian processes (GaP), which resemble kernel-based methods for regression.
However, GaPs develop these kernels adaptively based on the available data. They also provide probability distributions for the respective model parameters.
GaPs have been used extensively in problems related to time-dependent problems and they may be considered competitors, albeit more costly, to RNNs and echo state networks. Finally, we note the use of GaPs as surrogates for expensive cost 

%%%%%%%%%%%%%%%%%%%%%%
%%% MODELING
%%%%%%%%%%%%%%%%%%%%%%
\section{FLOW MODELING WITH MACHINE LEARNING}\label{sec:modeling}
 
First principles, such as  conservation laws, have been the dominant building blocks for flow modeling over the past centuries.  However, for high Reynolds numbers, scale resolving simulations using the most prominent model in fluid mechanics, the  Navier-Stokes equations,  is beyond our current computational resources.  An alternative is to perform simulations based on approximations of these equations  or  laboratory experiments for a specific configuration. 
However, simulations and experiments are expensive for iterative optimization, and simulations are often too slow for real-time control~\citep{Brunton2015amr}.
Consequently, considerable effort has been placed on obtaining accurate and efficient reduced-order models that capture essential flow mechanisms at a fraction of the cost~\citep{Rowley2016arfm}. 
Machine learning presents new avenues for dimensionality reduction and  reduced order modeling  in fluid mechanics by providing a concise framework  that complements and extends existing methodologies. 

\begin{marginnote}[]
\entry{Reduced-order Model (ROM)}{Representation of a high-dimensional system in terms of a low-dimensional one, balancing accuracy and efficiency.}
\end{marginnote}
We distinguish two complementary directions: dimensionality reduction and reduced-order modeling. 
Dimensionality reduction involves extracting key features and dominant patterns that may be used as reduced coordinates where the fluid is compactly and efficiently described~\citep{Taira2017aiaa}. 
Reduced-order modeling describes the spatiotemporal evolution of the flow  as a parametrized dynamical system, although it may also involve developing a statistical map from parameters to averaged quantities, such as drag. 

There have been significant efforts to identify coordinate transformations and reductions that simplify dynamics and capture essential flow physics:  the proper orthogonal decomposition (POD) is a notable example ~\citep{Lumley1970book}. 
Model reduction, such as Galerkin projection of the Navier-Stokes equations onto an orthogonal basis of POD modes, benefits from a close connection to the governing equations; however, it is intrusive, requiring human expertise to develop models from a working simulation. 
Machine learning 
% , entails algorithms such as the POD, and  
constitutes a rapidly growing body of modular algorithms for data-driven system identification and modeling. Unique aspects of data-driven modeling of fluid flows include the availability of partial prior knowledge of the governing equations, constraints, and symmetries. With advances in simulation capabilities and experimental techniques, fluid dynamics is becoming a data rich field, thus amenable to a wealth of machine learning algorithms. 

In this review, we distinguish machine learning algorithms to model flow  1)  \emph{kinematics} through the extraction flow features and 2)  \emph{dynamics} through the adoption of various learning architectures.

\subsection{Flow Feature Extraction}
Pattern recognition and data mining are core strengths of machine learning with many  techniques that are readily applicable to spatiotemporal flow data. 
We distinguish linear and nonlinear dimensionality reduction techniques, followed by clustering and classification. 
We also consider accelerated measurement and computation strategies, as well as methods to process experimental flow field data.

\subsubsection{Dimensionality reduction: Linear and nonlinear embeddings}
A common approach in fluid dynamics simulation and modeling is to define an orthogonal linear transformation from physical coordinates into a \emph{modal} basis. 
The POD provides such an orthogonal basis for complex geometries based on empirical measurements. 
\cite{Sirovich:1987} introduced the snapshot POD, which reduces the computation to a simple data-driven procedure involving a singular value decomposition. 
Interestingly, in the same year, Sirovich used POD to generate a low-dimensional feature space for the classification of human faces, which is a foundation for much of modern computer vision~\citep{Sirovich:1987b}.

\begin{figure}
\begin{center}
\vspace{-.2in}
\begin{overpic}[width=1.0\textwidth]{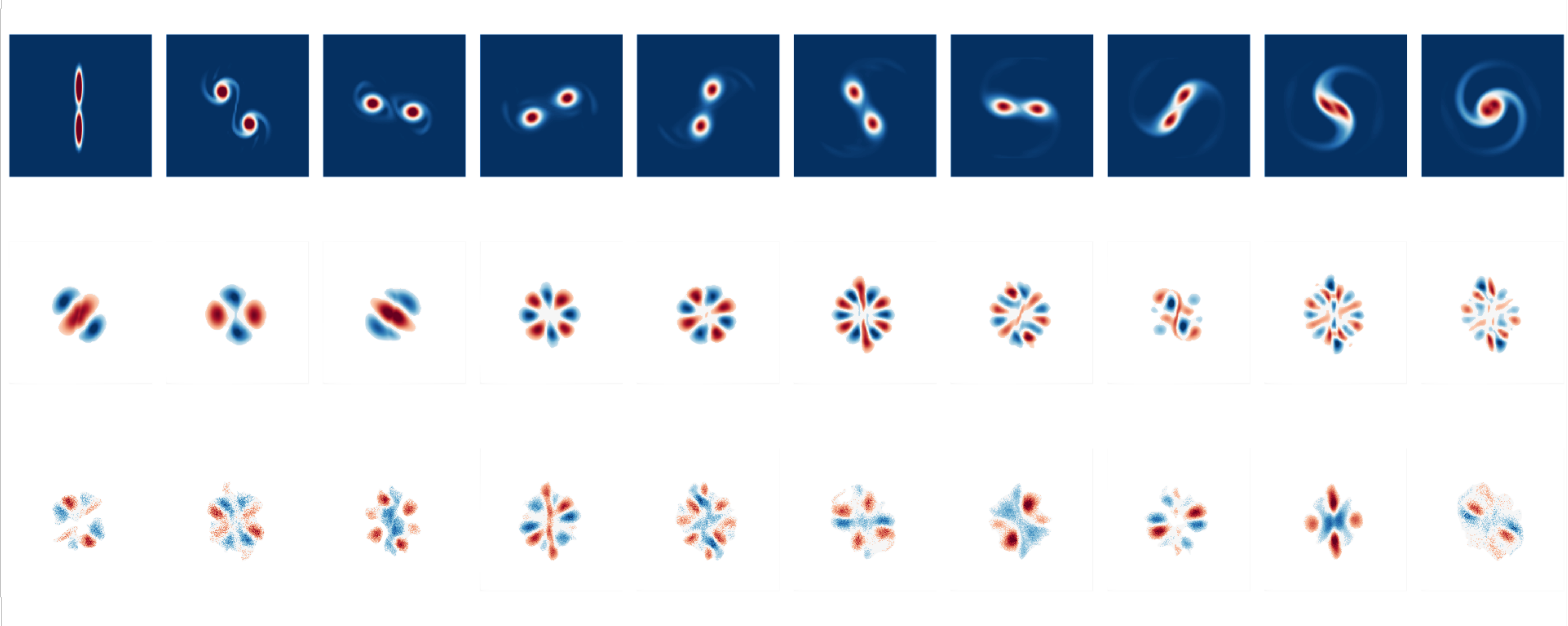}
\put(4,39){Flow snapshots}
\put(4,23){POD modes}
\put(4,11){Autoencoder modes}
\end{overpic}
\vspace{-.1in}
\caption{{\it Unsupervised learning example}: Merging of two vortices (top), POD modes (middle) and respective modes from a linear auto-encoder (bottom). Note that unlike POD modes, the autoencoder modes are not orthogonal. 
}\label{LinearAE}
\end{center}
\vspace{-.175in}
\end{figure}

POD is closely related to the algorithm of principal component analysis (PCA), one of the fundamental algorithms of applied statistics and  machine learning, to describe correlations in high-dimensional data.  
We recall that the PCA can be expressed as a two layer neural network, called an autoencoder, to compress high-dimensional data for a compact representation as shown in Fig.~\ref{NNautoencoder}. 
This network embeds high-dimensional data into a low-dimensional latent space, and then decodes from the latent space back to the original high-dimensional space. 
When the network nodes are linear and the encoder and decoder are constrained to be transposes of one another, the autoencoder is closely related to the standard POD/PCA decomposition (~\citep{baldi1989neural}, please see also Fig.~\ref{LinearAE}).
However, the structure of the neural network autoencoder is modular, and by using  nonlinear activation units for the nodes, it is possible to develop \emph{nonlinear} embeddings, potentially providing more compact coordinates. This observation led to the development of one of the first applications of deep neural networks to reconstruct the near wall velocity field in a turbulent channel flow using wall pressure and shear~\citep{Milano2002jcp}. More powerful autoencoders are today available in the ML community and this link deserves further exploration.

On the basis of the universal approximation theorem~\citep{hornik1989multilayer},  stating  that a sufficiently large neural network can represent an arbitrarily complex input--output function, \emph{deep} neural networks are increasingly used to obtain more effective nonlinear coordinates for complex flows. 
However, deep learning  often implies the availability of large volumes of training data that far exceed the parameters of the network. The resulting models are usually good for interpolation but may not be suitable for extrapolation when the new input data  have different probability distributions than the training data (see Eq.~\eqref{Eq:RiskFunctional}).  
In many modern machine learning applications, such as image classification, the training data are so vast that it is natural to expect  that most future classification tasks will fall within an interpolation of the training data. 
For example, the ImageNet data set in 2012~\citep{Krizhevsky2012nips} contained over 15 million labeled images, which sparked the current movement in \emph{deep learning}~\citep{Lecun2015nature}. 
Despite the abundance of data from experiments and simulations the fluid mechanics community is still distanced from this working paradigm.  
However, it may be possible in the coming years to curate large, labeled  and complete enough fluid databases to facilitate the deployment of such deep learning algorithms. 

\subsubsection{Clustering and classification}
Clustering and classification are cornerstones of machine learning. There are  dozens of mature algorithms to choose from, depending on the size of the data and the desired number of categories. 
The $k$-means algorithm has been successfully employed by \cite{Kaiser2014jfm} to develop a data-driven discretization of a high-dimensional phase space for the fluid mixing layer. 
This low-dimensional representation, in terms of a small number of clusters, enabled tractable Markov transition models for how the flow evolves in time from one state to another. 
Because the cluster centroids exist in the data space, it is possible to associate each cluster centroid with a physical flow field, lending additional interpretability. 
In \cite{Amsallem2012nme} $k$-means clustering was used to partition phase space into separate regions, in which local reduced-order bases were constructed, resulting in improved stability and robustness to parameter variations. 

Classification is also widely used in fluid dynamics to distinguish between various canonical behaviors and dynamic regimes. 
Classification is a supervised learning approach where labeled data is used to develop a model to sort new data into one of several categories. 
Recently, \cite{colvert2018classifying} investigated the classification of wake topology (e.g., 2S, 2P+2S, 2P+4S) behind a pitching airfoil from local vorticity measurements using neural networks; extensions have compared performance for various types of sensors~\citep{alsalman2018training}. 
In \cite{wang2017detecting} the $k$ nearest neighbors (KNN) algorithm was used to detect exotic wakes. 
Similarly, neural networks have been combined with dynamical systems models to detect flow disturbances and estimate their parameters~\citep{hou2019machine}.
Related graph and network approaches in fluids by \cite{nair2015network} have been used for community detection in wake flows~\citep{meena2018network}. 
Finally, one of the earliest examples of machine learning classification in fluid dynamics by \cite{Bright2013pof} was based on sparse representation~\citep{Wright2009ieeetpami}. 

\subsubsection{Sparse and randomized methods}
In parallel to machine learning, there have been great strides in sparse optimization and randomized linear algebra.  
Machine learning and sparse algorithms are synergistic, in that underlying low-dimensional representations facilitate sparse measurements~\citep{Manohar2017csm} and fast randomized computations~\citep{Halko2011siamreview}. 
Decreasing the amount of data to train and execute a model is important when a fast decision is required, as in control. In this context algorithms for the efficient acquisition and reconstruction of sparse signals, such as compressed sensing \cite{Donoho2006}, have already been leveraged for compact representations of wall-bounded turbulence~\citep{Bourguignon2014pof} and for POD based flow reconstruction~\citep{Bai2014aiaa}.

Low-dimensional structure in data also facilitates dramatically accelerated computations via randomized linear algebra~\citep{Mahoney2011,Halko2011siamreview}. 
If a matrix has low-rank structure, then there are extremely efficient matrix decomposition algorithms based on random sampling; this is closely related to the idea of sparsity and the high-dimensional geometry of sparse vectors. 
The basic idea is that if a large matrix has low-dimensional structure, then with high probability this structure will be preserved after projecting the columns or rows onto a random low-dimensional subspace, facilitating efficient downstream computations. 
These so-called \emph{randomized} numerical methods have the potential to transform computational linear algebra, providing accurate matrix decompositions at a fraction of the cost of deterministic methods. 
For example, randomized linear algebra may be used to efficiently compute the singular value decomposition, which is used to compute PCA~\citep{rokhlin2009randomized,Halko2011siamreview}. 

\subsubsection{Super resolution and flow cleansing}
Much of machine learning is focused on imaging science, providing robust approaches to improve resolution and remove noise and corruption based on statistical inference. 
These \emph{super resolution} and \emph{de-noising} algorithms have the potential to improve the quality of both simulations and experiments in fluids. 

Super resolution involves the inference of a high-resolution image from low-resolution measurements, leveraging the statistical structure of high-resolution training data. 
Several approaches have been developed for super resolution, for example based on a library of examples~\citep{Freeman2002ieeecga}, sparse representation in a library~\citep{Yang2010ieeetip}, and most recently based on convolutional neural networks~\citep{dong2014learning}. 
Experimental flow field measurements from particle image velocimetry (PIV)~\citep{Willert1991ef,Adrian1991arfm} provide a compelling application where there is a tension between local flow resolution and the size of the imaging domain.
Super resolution could leverage expensive and high-resolution data on smaller domains to improve the resolution on a larger imaging domain.
Large eddy simulations (LES)~\citep{germano1991dynamic,meneveau2000scale} may also benefit from super resolution to infer the high-resolution structure inside a low-resolution cell that is required to compute boundary conditions. 
Recently \cite{fukami2018super} have developed a CNN-based super-resolution algorithm and demonstrated its effectiveness on turbulent flow reconstruction, showing that the energy spectrum is accurately preserved. 
One drawback of super-resolution is that it is often extremely costly computationally, making it useful for applications where high-resolution imaging may be prohibitively expensive; however, improved neural-network based approaches may drive the cost down significantly. We note also that \cite{xie2018tempogan} recently employed GANs for super-resolution.

The processing of experimental PIV and particle tracking has been also one of the first applications of machine learning.
Neural networks have been used for fast PIV~\citep{Knaak1997nn} and particle tracking velocimetry~\citep{Labonte1999ef}, with impressive demonstrations for three-dimensional Lagrangian particle tracking~\citep{Ouellette2006ef}. 
More recently, deep convolutional neural networks have been used to construct velocity fields from PIV image pairs~\citep{lee2017piv}. 
Related approaches have also been used to detect spurious vectors in PIV data~\citep{Liang2003ef} to remove outliers and fill in corrupt pixels.

\subsection{Modeling Flow Dynamics}
A central goal of modeling is to balance efficiency and accuracy. When modeling physical systems, interpretability and generalizability are also critical considerations. 

\subsubsection{Linear models through nonlinear embeddings: DMD and Koopman analysis}
Many classical techniques in system identification may be considered machine learning, as they are data-driven models that generalize beyond the training data. 
% %
The dynamic mode decomposition (DMD)~\citep{Schmid2010jfm,Kutz2016book} is a modern approach, to extract spatiotemporal coherent structures from time-series data of fluid flows, resulting in a low-dimensional linear model for the evolution of these dominant coherent structures. 
DMD is based on data-driven regression and is equally valid for time-resolved experimental and numerical data. 
DMD is closely related to the Koopman operator~\citep{Rowley2009jfm,Mezic2013arfm}, which is an infinite dimensional linear operator that describes how \emph{all} measurement functions of the system evolve in time.
Because the DMD algorithm is based on linear flow field measurements (i.e., direct measurements of the fluid velocity or vorticity field), the resulting models may not be able to capture nonlinear transients. 

Recently, there has been a concerted effort to identify \emph{nonlinear} measurements that evolve linearly in time, establishing a coordinate system where the nonlinear dynamics appear linear. 
The extended DMD~\citep{Williams2015jcd} and variational approach of conformation dynamics (VAC)~\citep{noe2013variational,nuske2016variational} enrich the model with nonlinear measurements, leveraging kernel methods~\citep{Williams2015jcd} and dictionary learning~\citep{Li2017chaos}. 
These special nonlinear measurements are generally challenging to represent, and deep learning architectures are now used to identify nonlinear Koopman coordinate systems where the dynamics appear linear~\citep{Wehmeyer2018jcp,Mardt2018natcomm,Takeishi2017nips,Lusch2018natcomm}. 
The VAMPnet architecture~\citep{Wehmeyer2018jcp,Mardt2018natcomm} uses a time-lagged auto-encoder and a custom variational score to identify Koopman coordinates on an impressive protein folding example. 
Based on the performance of VAMPnet, fluid dynamics may benefit from neighboring fields, such as molecular dynamics, which have similar modeling issues, including stochasticity, coarse-grained dynamics, and massive separation of time scales. 

\subsubsection{Neural network modeling}
Over the last three decades neural networks have been used to model dynamical systems and fluid mechanics problems.
Early examples include the use of NNs to learn the solutions of  ordinary and partial differential equations~\citep{dissanayake1994neural,gonzalez1998identification,lagaris1998artificial}. We note that the potential of this work has not been fully explored and in recent years there is further  advances   \citep{Chen2018,raissi2018hidden} including discrete and continuous in time networks. We note also the possibility of using these methods to uncover latent variables and reduce the number of parametric studies often associated with partial differential equations \cite{raissi2019physics}. 
Neural networks are also frequently employed in nonlinear system identification techniques, such as NARMAX, which are often used to model fluid systems~\citep{Glaz2010aiaa}. 
In fluid mechanics, neural networks were widely used to model heat transfer~\citep{Jambunathan1996ijhmt}, turbomachinery~\citep{Pierret1998asme}, turbulent flows~\citep{Milano2002jcp}, and other problems in aeronautics~\citep{Faller1996pas}. 

Recurrent Neural Netwosk with LSTMs (\cite{hochreiter1997long} have been revolutionary for speech recognition, and they are considered one of the landmark successes of artificial intellignece.  The are currently being used to model dynamical systems and for data driven predictions of extreme events~\citep{Wan2018,vlachas2018data}. An interesting finding of these studies is that combining data driven and reduced order models is a potent method that outperforms each of its components on a number of studies.
Generative adversarial networks (GANs)~\citep{goodfellow2014generative} are also being used to infer dynamical systems from data~\citep{wu2018deep}. 
GANs have potential to aid in the modeling and simulation of turbulence~\citep{kim2018deep}, although this field is nascent, yet well worthy of exploration.

Despite the promise and widespread use of neural networks in dynamical systems, a number of challenges remains. 
Neural networks are fundamentally \emph{interpolative}, and so the function is only well approximated in the span (or under the probability distribution) of the sampled data  used to train them. Thus, caution should be exercised when using neural network models for an extrapolation task. 
In many computer vision and speech recognition examples, the training data are so vast that nearly all future tasks may be viewed as an interpolation on the training data. However,  this scale of training has not been achieve to date in fluid mechanics. 
Similarly, neural network models are prone to overfitting, and care must be taken to cross-validate models on a sufficiently chosen test set; best practices are discussed in \cite{Goodfellow-et-al-2016}. 
Finally, it is important to explicitly incorporate physical properties such as symmetries, constraints, and conserved quantities. 

\subsubsection{Parsimonious nonlinear models}
Parsimony is a recurring theme in mathematical physics, from Hamilton's principle of least action to the apparent simplicity of many governing equations.  
In contrast to the raw representational power of neural networks, machine learning algorithms are also being employed to identify minimal models that balance predictive accuracy with model complexity, preventing overfitting and promoting interpretability and generalizability. 
Genetic programming has  been used to discover conservation laws and governing equations~\citep{Schmidt2009science}. 
Sparse regression in a library of candidate models has also been proposed to identify dynamical systems~\citep{Brunton2016pnas} and partial differential equations~\citep{Rudy2017sciadv,Schaeffer2017prsa}. 
\cite{Loiseau2017jfm} identified sparse reduced-order models of several flow systems, enforcing energy conservation as a constraint. 
In both genetic programming and sparse identification, a Pareto analysis is used to identify models that have the best tradeoff between model complexity, measured in number of terms, and predictive accuracy. 
In cases where the physics is known, this approach typically discovers the correct governing equations, providing exceptional generalizability compared with other leading algorithms in machine learning.

\subsubsection{Closure models with machine learning}
The use of machine learning to develop turbulence closures is an active area of research~\citep{Duraisamy2018arfm}. 
The extremely wide range of spatiotemporal scales in turbulent flows makes it exceedingly costly to resolve all scales in simulation, and even with Moore's law, we are likely several decades away from resolving all scales in configurations of industrial interest (e.g., aircraft turbines, submarines, etc.). 
It is common to truncate small scales and model their effect on the large scales with a \emph{closure model}. 
Common approaches include Reynolds averaged Navier Stokes (RANS) and large eddy simulation (LES). 
However, these models may require careful tuning to match fully resolved simulations or experiments.

\begin{figure}
\begin{center}
\vspace{-.25in}
\includegraphics[width=.85\textwidth]{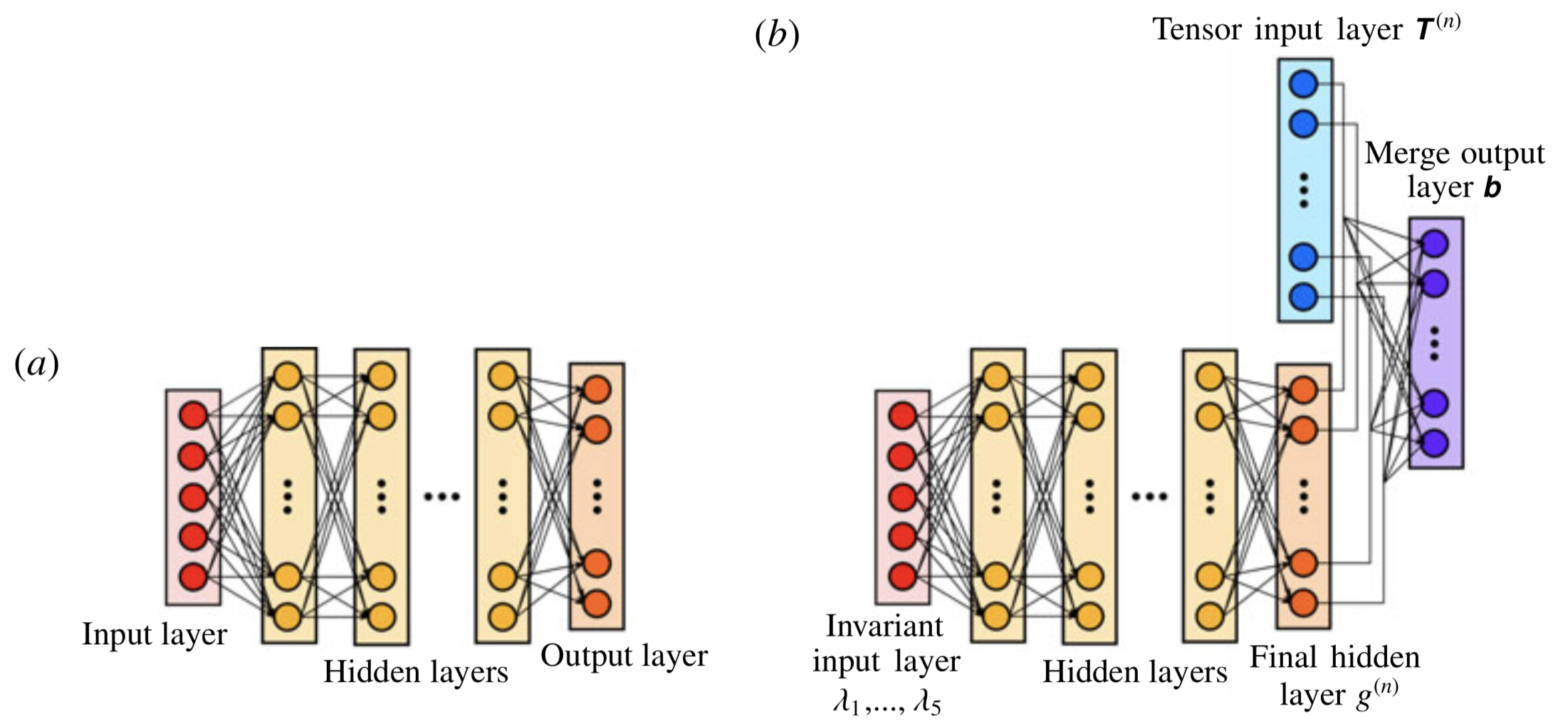}
\vspace{-.2in}
\caption{Comparison of standard neural network architecture (a) with modified neural network for identifying Galilean invariant Reynolds stress models (b), reproduced from~\cite{Ling2016jfm} with permission from {\it The Journal of Fluid Mechanics}. Symbols: 
 $ \lambda_1 , \cdots , \lambda_5$: five tensor invariants
  $T^(n)$: isotropic basis tensors
  $g^(n)$: scalar coefficients weighing the basis tensors
  $b$: anisotropy tensor}\label{Ling2016jfm}
\end{center}
\vspace{-.2in}
\end{figure}

Machine learning has been used to identify and model discrepancies in the Reynolds stress tensor between a RANS model and high-fidelity simulations~\citep{Ling2015pf,Parish2016jcp,Ling2016jfm,Xiao2016jcp,Singh2017aiaaj,Wang2017prf}. 
\cite{Ling2015pf} compare support vector machines, Adaboost decision trees, and random forests to classify and predict regions of high uncertainty in the Reynolds stress tensor. 
\cite{Wang2017prf} use random forests to built a supervised model for the discrepancy in the Reynolds stress tensor. 
\cite{Xiao2016jcp} leveraged sparse online velocity measurements in a Bayesian framework to infer these discrepancies. 
In related work, \cite{Parish2016jcp} develop the \emph{field inversion and machine learning} modeling framework, that builds corrective models based on inverse modeling.
This framework was later used by \cite{Singh2017aiaaj} to develop a neural network enhanced correction to the Spalart-Allmaras RANS model, with excellent performance. 
A key result by \cite{Ling2016jfm} employed the first \emph{deep} network architecture with many hidden layers to model the anisotropic Reynolds stress tensor, as shown in Fig.~\ref{Ling2016jfm}.
Their novel architecture incorporates a multiplicative layer to embed Galilean invariance into the tensor predictions. 
This provides an innovative and simple approach to embed known physical symmetries and invariances into the learning architecture~\citep{Ling2016jcp}, which we believe will be essential in future efforts that combine learning for physics. 
For large eddy simulation closures, \cite{Maulik2019jfm} have employed artificial neural networks to predict the turbulence source term from coarsely resolved quantities. 

\subsubsection{Challenges of machine learning for dynamical systems}
Applying machine learning to model physical dynamical systems poses a number of unique challenges and opportunities. 
Model interpretability and generalizability are essential cornerstones in physics. A well crafted model will yield hypotheses for new phenomena that have not been observed before. 
This principle is clearly exhibited in the parsimonious formulation of classical mechanics in Newton's second law. 

High-dimensional systems, such as those encountered in unsteady fluid dynamics,  have the  challenges of multi-scale dynamics, sensitivity to  noise and disturbances, latent variables and transients, all of which require careful attention when applying machine learning techniques. 
In machine learning for dynamics, we distinguish two tasks:  discovering unknown physics and improving models by incorporating  known physics. 
Many learning architectures, cannot readily incorporate physical constraints in the form of symmetries, boundary conditions, and global conservation laws. 
This is a critical area for continued development and  a number of recent works have presented 
 generalizable physics models~\citep{battaglia2018relational}. 

%%%%%%%%
%%%%%%%%
%%%%%%%%
\section{FLOW OPTIMIZATION AND CONTROL USING MACHINE LEARNING}\label{sec:control}
Learning algorithms are well suited to flow  optimization  and control problems
involving ``black-box" or multimodal cost functions. 
These algorithms are iterative and often require several orders of magnitude more cost function evaluations  than gradient based algorithms  ~\citep{Bewley2001jfm}. Moreover they do not offer guarantees of convergence and we suggest that they are avoided when techniques such as adjoint methods are applicable. 
At the same time, techniques such as reinforcement learning have been shown to outperform even optimal flow control strategies~\citep{Novati2019}. 
Indeed there are several classes of flow control and optimization problems where learning algorithms may be the method of choice as described below.

In contrast to flow modeling, learning algorithms for optimization and control interact with the data sampling process in several ways. First, in line with the modeling efforts described in earlier sections, 
machine learning can be applied to develop explicit surrogate models that relate the cost function and the control/optimization parameters. Surrogate models such as neural networks can then be amenable even to gradient based methods, although they often get stuck in local minima. Multi-fidelity algorithms \citep{Perdikaris2016} can also be employed to combine surrogates with the cost function of the complete problem. As the learning progresses, new data are requested as guided by the results of the optimization.   Alternatively, the optimization or control problem may be described in terms of learning probability distributions of parameters that minimize the cost function. These probability distributions are constructed from cost function samples obtained during the optimization process.  
Furthermore, the high-dimensional and non-convex optimization procedures that are currently employed to train nonlinear learning machines are well-suited to the high-dimensional, nonlinear optimization problems in flow control.  

We remark that the lines between optimization and control are becoming blurred by the availability of powerful computers (see focus box).   However, the range of critical spatiotemporal scales and the non-linearity of the underlying processes will likely render real-time optimization for flow control a challenge for decades to come. 

%%%%%%%%%%%%%%%%%%%%%%
%%% OPTIMIZING AND CONTROLLING
%%%%%%%%%%%%%%%%%%%%%%
\begin{textbox}[b]\section{OPTIMIZATION AND CONTROL: BOUNDARIES ERASED BY FAST COMPUTERS}
Optimization and control are intimately related, and the boundaries are becoming even less distinct with increasingly fast computers, as summarized in~\cite{tsiotras2017toward} (page 195):

``{\em 
Interestingly, the distinction between optimization and control is largely semantic and (alas!) implementation-dependent. If one has the capability of solving optimization problems fast enough on the fly to close the loop, then one has (in principle) a feedback control law...
Not surprisingly then, the same algorithm can be viewed as solving an optimization or a control problem, based solely on the capabilities of the available hardware. With the continued advent of faster and more capable computer hardware architectures, the boundary between optimization and control will become even more blurred. However,
when optimization is embedded in the implementation of feedback
control, the classical problems of control such as robustness to
model uncertainty, time delays, and process and measurement noise
become of paramount importance, particularly for high-performance
aerospace systems."} 
\end{textbox}

\subsection{Stochastic Flow Optimization: Learning Probability Distributions} 
Stochastic optimization includes evolutionary strategies and genetic algorithms, which were originally developed based on bio-inspired principles. However, over the last 20 years these algorithms have been placed in a learning framework~\citep{Kern:2004}.

 Stochastic optimization has found widespread use in engineering design, in particular as many engineering problems involve ``black-box" type of cost functions. A much abbreviated list of applications include aerodynamic shape optimization~\citep{Giannakoglou2006}, uninhabited aerial vehicles (UAVs) ~\citep{Hamdaoui2010ja}, shape and motion optimization in artificial swimmers~\citep{Gazzola2012,VanRees2015}, and improved power extraction in crossflow turbines~\citep{Strom2017natenergy}. 
 We refer to the review article by \cite{Skinner2018} for an extensive comparison of gradient-based and stochastic optimization algorithms for aerodynamics.
 
 These algorithm involve large numbers of iterations, and they can benefit from massively parallel computer architectures. Advances in automation have also facilitated their application in experimental~\citep{Strom2017natenergy,Martin2018} and industrial settings~\citep{Bueche:2002}. 
We note that stochastic optimization algorithms are well-suited to address the experimental and industrial challenges associated with uncertainty, such as unexpected system behavior, partial descriptions of the system and its environment, and exogenous disturbances. 
\cite{Hansen2009ieeetec} proposed an approach to enhance the capabilities of evolutionary algorithms for online optimization of a combustor test-rig.

Stochastic flow optimization will continue to benefit from advances in computer hardware and experimental techniques.  At the same time, convergence proofs, explainability, and reliability are outstanding issues that need to be taken into consideration when deploying such algorithms in fluid mechanics problems. Hybrid algorithms, combining in a problem specific manner stochastic techniques and gradient-based methods may offer the best strategy for flow control problems.

%%%%%%%%%%%
%%%%%%%%%%% CONTROL
%%%%%%%%%%%
\subsection{Flow Control with Machine Learning} 
Feedback flow control modifies the behavior of a fluid dynamic system through actuation that is informed by sensor measurements.  
Feedback is necessary to stabilize an unstable system, attenuate sensor noise, and compensate for external disturbances and model uncertainty.  
Challenges of flow control include a high-dimensional state, nonlinearity, latent variables, and time delays.  
Machine learning algorithms have been used extensively in control, system identification, and sensor placement.

%%% NEURAL NETS IN OPPOSITION CONTROL
\subsubsection{Neural networks for control}

Neural networks have received significant attention  for system identification (see Sec.~\ref{sec:modeling}) and control, including applications in aerodynamics~\citep{Phan1994}.
The application of NNs to turbulence flow control was pioneered in~\cite{Lee1997pof}. 
The skin-friction drag of a turbulent boundary layer was reduced using local wall-normal blowing and suction based on few skin friction sensors. A sensor-based control law was learned from a known optimal full-information controller, with little loss in overall performance. Furthermore, a single-layer network was optimized for skin-friction drag reduction without incorporating any prior knowledge of the actuation commands. 
Both strategies led to a conceptually simple local opposition control.
Several other studies employ neural networks, e.g. for phasor control \citep{Rabault2019jfm} 
or even frequency cross talk. % \citep{Rabault2015private}.
The price for the theoretical advantage of approximating arbitrary nonlinear control laws is the need for many parameters to be optimized. 
Neural network control may require exorbitant computational or experimental resources for configurations with complex high-dimensional nonlinearities and many sensors and actuators. 
At the same time, the training time of neural networks has been improved by several orders of magnitude since these early applications, which  warrants further investigation into their potential for flow control.

\subsubsection{Genetic algorithms for control}

Genetic algorithms have been deployed to solve a number of flow control problems. They require that the structure of the control law is pre-specified and contains only a few adjustable parameters. 
An example of GA for control design in fluids was used for experimental mixing optimization 
of the backward-facing step~\citep{Benard2016ef}. 
As with neural network control, the learning time increases with the number of parameters, making it challenging or even prohibitive for controllers with nonlinearities, e.g.\ a constant-linear-quadratic law, with signal history, e.g.\ a Kalman filter, or with multiple sensors and actuators. 

Genetic programming has been used extensively in active control for engineering applications~\citep{Dracopoulos1997book,Fleming2002cep} and in recent years in several flow control plants.
This includes the learning of multi-frequency open-loop actuation, 
multi-input sensor feedback,
and distributed control. 
We refer to~\cite{Duriez2016book} for an in-depth description of the method
and to~\cite{Noack2018fssic} for an overview of the plants.
We remark that most control laws have been obtained within 1000 test evalutations, each requiring only few seconds in a wind-tunnel.

\subsection{Flow Control via Reinforcement Learning}

\begin{figure}
  \centering
  \vspace{-.4in}
  \includegraphics[width=1.175\textwidth]{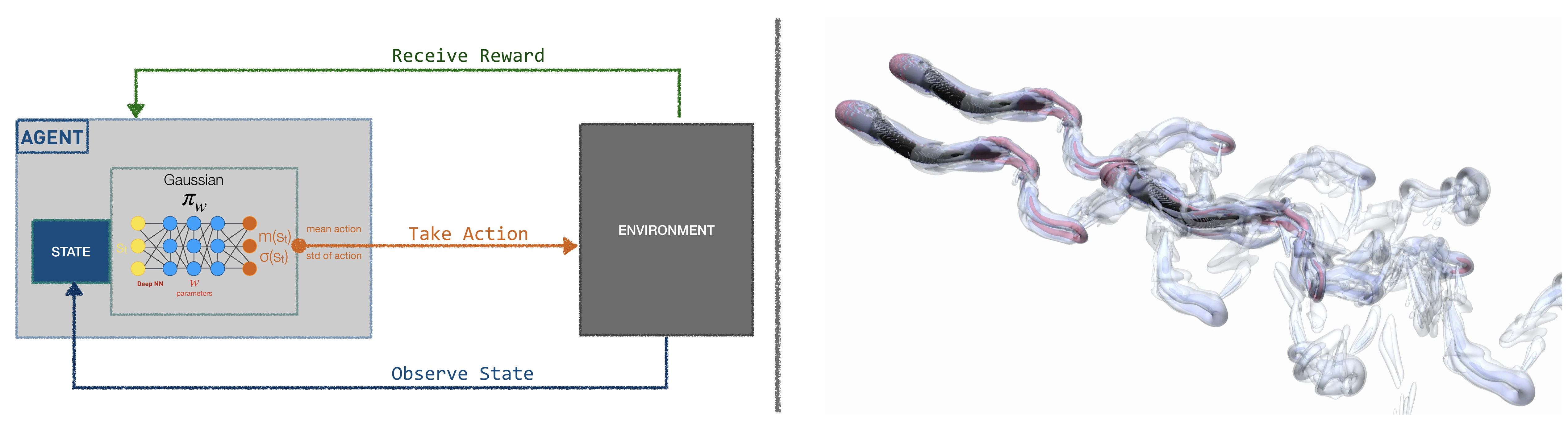}
  \vspace{-0.1in}
  \caption{Deep reinforcement learning schematic (left), and application to the study of the collective motion of fish via the Navier-Stokes equations (right; ~\cite{Verma2018EfficientLearning}).Symbols:
  $S_t$:state,
  $\pi_w$:policy,
  $W$:parameters, 
  $m(S_t),\sigma(S_t)$:mean, standard  deviation for action}
  \label{fig:appsDeepRL}
  \vspace{-0.1in}
\end{figure}
In recent years RL has advanced beyond the realm of games and has become a fundamental mode of problem solving in a growing number of domains, including to reproduce the dynamics of hydrological systems~\citep{Loucks2005}, actively control the oscillatory laminar flow around bluff bodies~\citep{Gueniat2016}, study the individual~\citep{gazzola2014reinforcement} or the collective motion of fish~\citep{Gazzola2016LearningInteractions,Novati2017,Verma2018EfficientLearning}, maximize the range of simulated~\citep{Reddy2016LearningEnvironments} and robotic~\citep{Reddy19} gliders, optimize the kinematic motion of UAVs~\citep{Kim2004nips,Tedrake2009isrr}, and  optimize the motion of microswimmers~\citep{Colabrese2017,Colabrese2018}. 
Figure~\ref{fig:appsDeepRL} provides a schematic of reinforcement learning with an examples showcasing its application to collective swimming as resolved by the Navier-Stokes equations.  

\begin{marginnote}
\entry{Reinforcement Learning}{An agent learns a policy of actions that maximize its long term rewards by interacting with its environment.}
\end{marginnote}
Fluid mechanics knowledge is essential for applications of RL, as success or failure hinges on properly selecting states, actions, and rewards that reflect the governing mechanisms of the flow problem.  
Natural organisms and their sensors, such as the visual system in a bird or the lateral line in a fish, can guide the choice of states. 
As sensor technologies progress at a rapid pace, the algorithmic challenge may be that of optimal sensor placement~\citep{Papadimitriou2015,Manohar2017csm}.
The actions reflect the flow actuation device and may involve body deformation or wing flapping.  Rewards may include energetic factors, such as the cost of transport, or proximity to the center of a fish school to avoid predation. 
The computational cost of RL remains a challenge to its widespread adoption, but we believe this deficiency can be mediated by the parallelism inherent to RL. 
There is growing interest in methods designed to be transferable from low-accuracy (e.g. 2-dimensional) to high-accuracy (e.g. 3-dimensional) simulations~\citep{Verma2018EfficientLearning}, or from simulations to related real-world applications~\citep{richter2016}. 

%%%%%%%%%%%%%%%%%%%%%%
%%% OUTLOOK
%%%%%%%%%%%%%%%%%%%%%%
\section{DISCUSSION AND OUTLOOK}\label{sec:outlook} 
This review presents machine learning algorithms that could augment existing efforts for the study, modeling and control of fluid mechanics. The interface of the two fields has a long history and has attracted a renewed interest in the last few years.  The review addresses applications of machine learning in problems of flow modeling, optimization, and control in experiments and simulations.
It highlights  some successes of machine learning in critical fluid mechanics tasks, such as dimensionality reduction, feature extraction, PIV processing, super-resolution, reduced-order modeling, turbulence closure, shape optimization, and flow control.  It  discusses lessons learned from these efforts and justify the current interest in light of the technological advances of our times. 
Machine learning comprises data-driven optimization and applied regression techniques that are well-suited for high-dimensional, nonlinear problems, such as those encountered in fluid dynamics; fluid mechanics expertise will be necessary to formulate these optimization and regression problems. 

Machine learning algorithms present an arsenal of tools,  largely unexplored in fluid mechanics research, that can augment existing modes of inquiry. Fluid mechanics knowledge and centuries old principles such as conservation laws remain relevant in the era of big data. Such knowledge can help frame more precise questions and assist in reducing the large computational cost often associated with the application of  machine learning algorithms in flow control and optimization. The exploration and visualization of high-dimensional search spaces will be dramatically simplified by machine learning and increasingly capable high-performance computing resources. 

In the near future, experience with machine learning algorithms will help frame new questions in fluid mechanics, extending decades old linearized models and linear approaches to the nonlinear regime. The transition to the nonlinear realm of machine learning is facilitated  by the abundance of open source software and methods, and the prevalent openness of the ML community. In the long term machine learning will undoubtedly offer  a fresh look into old problems of fluid mechanics under the light of data. 
Interpreting the machine learning solutions, and refining the problem statement, will again require fluid mechanics expertise.  

A word of caution is necessary to balance the current excitement about data-driven research and the (almost magical) powers of machine learning. After all a machine learning algorithm will always provide some kind of answer to any question, that is based on its training data; data that may not be even relevant to the question at hand. Properly formulating the question, selecting the data as well as the learning machine and its training is a process where all parts are critical.
Applying machine learning algorithms to fluid mecahnics is faced with numerous outstanding challenges (and opportunities!). 
Although many fields of machine learning are concerned with raw predictive performance, applications in fluid mechanics often require models that are explainable, generalizable, and have guarantees. 

Although deep learning will undoubtedly become a critical tool in several aspects of flow modeling, not all machine learning is deep learning. 
It is important to consider several factors when choosing methods, including the quality and quantity of data, the desired inputs and outputs, the cost function to be optimized, whether or not the task involves interpolation or extrapolation, and how important it is for models to be explainable.  It is important to cross-validate machine learned models, otherwise results may be prone to overfitting. 
It is also important to develop and adapt machine learning algorithms that are not only \emph{physics informed} but also {\it physics consistent}, a major outstanding challenge in artificial intelligence. 
This review concludes  with a call for  action in the fluid mechanics community to further embrace open and reproducible research products and standards.
Reproducibility  is a cornerstone of science and a number of frameworks are currently developed to render this intop a systematic scientifc process (\cite{BarberCandes}). It is increasingly possible to document procedures, archive code, and host data so that others can reproduce results.  Data is essential for machine learning; thus, creating and curating \emph{benchmark} datasets and software will spur interest among researchers in related fields, driving progress. 
These fluid benchmarks are more challenging than the ``traditional" image data sets encountered in machine learning: fluids data is multi-modal and multi-fidelity; it has high-resolution in some dimensions and is sparse in others; many tasks balance multiple objectives; and foremost, our data comes from a dynamical system, where many tasks do not admit \emph{post-mortem} analysis. 

We are entering a new era in fluid mechanics research. 
Centuries of theoretical developments based on first principles are merging with data-driven analysis. This fusion could provide solutions to many long-sought problems in fluid dynamics and offers new hope for  enhanced understanding of turbulence and its governing mechanisms.
\begin{marginnote}
\entry{Reproducibility}{The process of documenting procedures, archiving code and data so that scientific results can be readily reproducible.}
\end{marginnote}

\begin{summary}[SUMMARY POINTS]
\vspace{-.075in}
\begin{enumerate}
%-----------------------------------------------------------------------
\item Machine learning entails powerful information processing algorithms that are relevant for modeling, optimization, and control of fluid flows. Effective problem solvers will have expertise in machine learning and in-depth knowledge of fluid mechanics.
\item Fluid mechanics has been traditionally concerned with big data.  
For decades it has used machine learning to understand, predict, optimize, and control flows. 
Currently, machine learning capabilities are advancing at an unprecedented rate, and fluid mechanics is beginning to tap into the full potential of these powerful methods.
%-----------------------------------------------------------------------
\item Many tasks in fluid mechanics, such as reduced-order modeling, shape optimization, and feedback control, may be posed as optimization and regression tasks. 
Machine learning can dramatically improve optimization performance and reduce convergence time. 
Machine learning is also used for dimensionality reduction, identifying low-dimensional manifolds and discrete flow regimes, which benefit understanding. 
%-----------------------------------------------------------------------
\item Flow control strategies have been traditionally based on the precise sequence: start with understanding, follow with modeling and then control.
The machine-learning paradigm suggests more flexibility in this sequence and iterations  between data driven and first principle approaches.
\end{enumerate}
\vspace{-.125in}
\end{summary}

\begin{issues}[FUTURE ISSUES]
\vspace{-.075in}
\begin{enumerate}%-----------------------------------------------------------------------
\item Machine learning algorithms often come without guarantees for performance, robustness, or convergence, even for well-defined tasks. 
How can interpretability, generalizability, and explainability of the results be achieved?
%-----------------------------------------------------------------------
\item Incorporating and enforcing known flow physics is a challenge and opportunity for machine learning algorithms. Can we hybridize effectively data driven and first principle approaches in fluid mechanics?
%-----------------------------------------------------------------------
\item There are many possibilities to discover new physical mechanisms, symmetries, constraints, and invariances from fluid mechanics  data. 

\item Data driven modeling can  be a potent alternative in  revisiting existing empirical laws in fluid mechanics.
%-----------------------------------------------------------------------
\item 
% Is there merit for open source repositories for machine learning software and related fluid mechanics data? 
Machine learning encourages open sharing of data and software. Can this assist the development of  frameworks for reproducible and open science in fluid mechanics?

\item 
% We  need a faster turnaround in evaluating machine learning related publications in fluid mechanics. 
Fluids researchers will benefit from interfacing with the machine learning community, where the latest advances are reported in peer reviewed conferences. 
\vspace{-.125in}
\end{enumerate}
\end{issues}

%Disclosure
\section*{DISCLOSURE STATEMENT}
The authors are not aware of any affiliations, memberships, funding, or financial holdings that
might be perceived as affecting the objectivity of this review. 

\vspace{-.2in}

% Acknowledgements
\section*{ACKNOWLEDGMENTS}
SLB acknowledges funding from the Army Research Office (ARO W911NF-17-1-0306, W911NF-17-1-0422) and the Air Force Office of Scientific Research (AFOSR FA9550-18-1-0200). 
BRN acknowledges funding 
by LIMSI-CNRS, Universit\'e Paris Sud (SMEMaG),
the French National Research Agency (ANR-11-IDEX-0003-02, ANR-17-ASTR-0022) 
and the German Research Foundation (CRC880, SE 2504/2-1, SE 2504/3-1).
PK acknowledges funding from the ERC Advanced Investigator Award (FMCoBe, No. 34117), the Swiss National Science Foundation and the Swiss Supercomputing center (CSCS).
We are grateful for discussions with Nathan Kutz (University of Washington), Jean-Christophe Loiseau (ENSAM ParisTech, Paris), Fran\c{c}ois Lusseyran (LIMSI-CNRS, Paris), Guido Novati (ETH Zurich), Luc Pastur (ENSTA ParisTech, Paris), and Pantelis Vlachas (ETH Zurich).

%\bibliographystyle{ar-style1}
%\bibliography{references} 

%%%% BIBLIOGRAPHY WITH HIGHLIGHTS

\end{document}